\documentclass[nonacm,sigconf,authordraft,review=false]{acmart}

\usepackage{natbib}
\setcitestyle{authoryear}
\usepackage{supertabular}
\usepackage{longtable}
\usepackage{tabularx}
\usepackage{pdflscape}
\usepackage{subcaption}
\usepackage{bbold}
\usepackage{algorithm}
\usepackage{algpseudocode}
\usepackage{xcolor}
\usepackage{graphicx}
\AtBeginDocument{%
  \providecommand\BibTeX{{%
    \normalfont B\kern-0.5em{\scshape i\kern-0.25em b}\kern-0.8em\TeX}}}

\copyrightyear{}
\acmYear{}
\acmDOI{}

\begin{document}

\title{Limit Order Book Dynamics and Order Size Modelling Using Compound Hawkes Process}

\author{Konark Jain}
\email{konark.jain.23@ucl.ac.uk}
\affiliation{%
  \institution{University College London; JP Morgan Chase \& Co.}
  \city{London}
  \country{UK}
}

\author{Nick Firoozye}
\email{n.firoozye@ucl.ac.uk}
\affiliation{%
  \institution{University College London}
  \city{London}
  \country{UK}
}

\author{Jonathan Kochems}
\email{jonathan.a.kochems@jpmorgan.com}
\affiliation{%
  \institution{JP Morgan Chase \& Co.}
  \city{London}
  \country{UK}
}

\author{Philip Treleaven}
\email{p.treleaven@ucl.ac.uk}
\affiliation{%
  \institution{University College London}
  \city{London}
  \country{UK}
}


\newcommand{\highlight}{} 
\begin{abstract}
Hawkes Process has been used to model Limit Order Book (LOB) dynamics in several ways in the literature however the focus has been limited to capturing the inter-event times while the order size is usually assumed to be constant. We propose a novel methodology of using Compound Hawkes Process for the LOB where each event has an order size sampled from a calibrated distribution. The process is formulated in a novel way such that the spread of the process always remains positive. Further, we condition the model parameters on time of day to support empirical observations. We make use of an enhanced non-parametric method to calibrate the Hawkes kernels and allow for inhibitory cross-excitation kernels. We showcase the results and quality of fits for an equity stock's LOB in the NASDAQ exchange and compare them against several baselines. Finally, we conduct a market impact study of the simulator and show the empirical observation of a concave market impact function is indeed replicated. 
\end{abstract}






\maketitle

\section{Introduction}

The Hawkes process, known for its high adaptability, offers a more comprehensive point process methodology for modeling order book arrivals than the Poisson process and its variants, without the need to explicitly model individual traders' behaviors in the market. Its capability to replicate microstructural details such as volatility clustering and correlated order flow makes it a suitable candidate for Limit Order Book (LOB) models. It is important to highlight that these point process models are mathematically descriptive, providing full transparency in their nature and thus are suitable for applications where black-box solutions are not preferred. In their comprehensive review and tutorial, \citealt{Bacry2015} outlay the major ideas of the Hawkes Process, its mathematical theory, some of its crucial properties and finally applications including a detailed review over the Order Book models. Recently, state dependent Hawkes Process have been quite popular (\citealt{pakannen2022},  \citealt{kirchner2022hawkes}, \citealt{mucciante2023estimation}, \citealt{wu2019queue}). However, as noted in \citealt{rambaldi2017role} and \citealt{lu2018order}, individual order's size is an important of aspect of the LOB which the Hawkes model alone is unable to capture. 

\subsection{Empirical Distributions of Order Sizes:}

\begin{figure}[h]
\begin{subfigure}{.5\columnwidth}
  \centering
  \includegraphics[width=.9\linewidth]{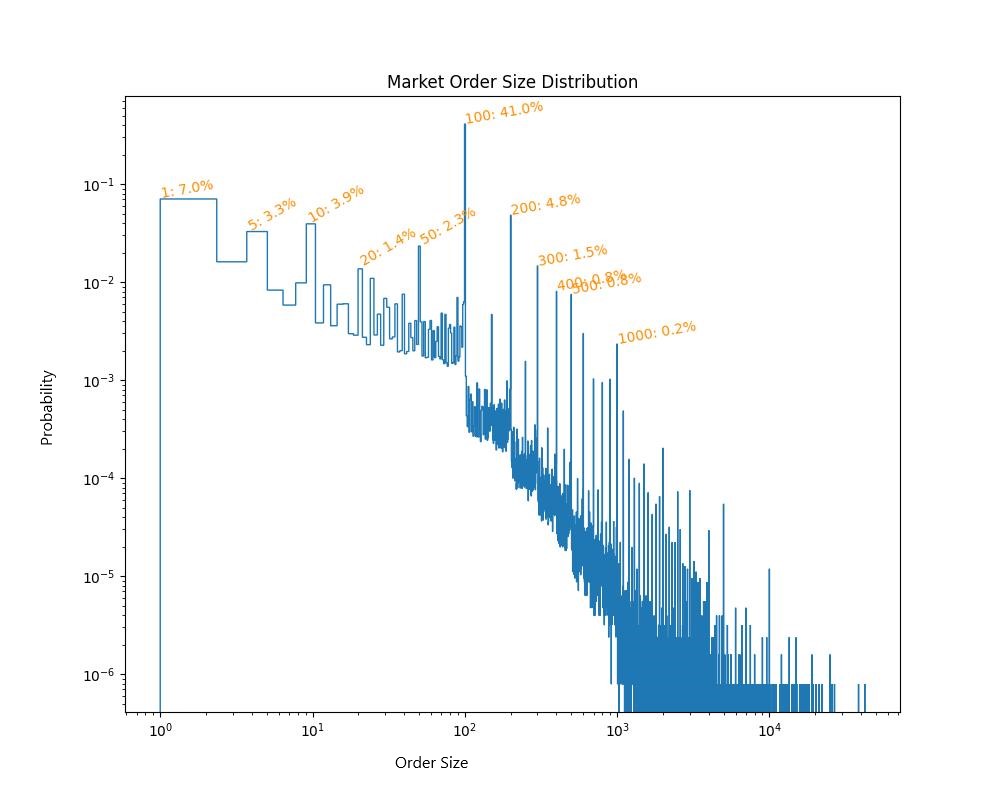}
  \caption{Market Orders' Size Distribution (log-log scale)}
  \label{fig:empMO}
\end{subfigure}%
\begin{subfigure}{.5\columnwidth}
  \centering
  \includegraphics[width=.9\linewidth]{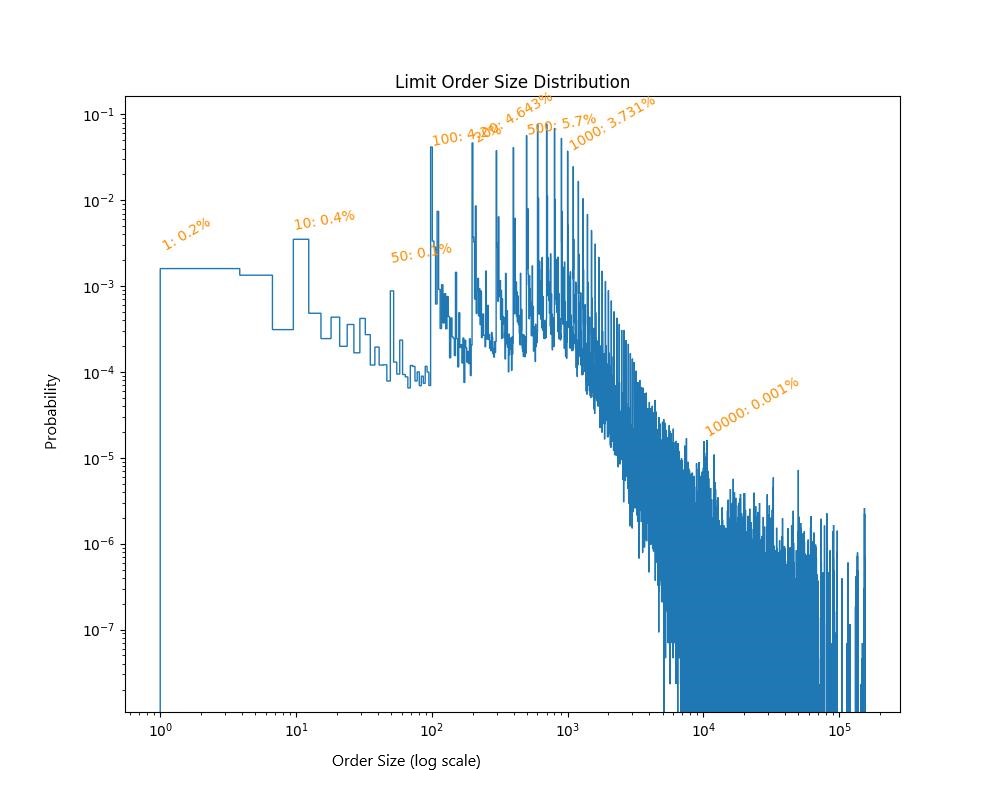}
  \caption{Limit Orders' Size Distribution (log-log scale)}
  \label{fig:empLO}
\end{subfigure}
\caption{Empirical Distribution of Order Sizes}
\label{fig:emp}
\end{figure}

We show here in Figure 1 two trading weeks' of data (10 dates) for Apple and plot the empirical histograms on a log-log scale. We only focus on top of the book cancels and limit orders. As Figures \ref{fig:empMO} and \ref{fig:empLO} show, the market orders and limit orders sizes have several spikes at round numbers indicating the preference of the traders. For example, we see more than 40\% of the market orders and 60\% of limit orders are of 100 size. Naturally, Cancel Orders' sizes are capped at the size of an individual order. In fact we observe that outright cancels (i.e. full quantity of an order is cancelled) constitute 99.3\% of all cancels.  

\subsection{Properties of the Order Book Dynamics:}
There have been several variations to the Hawkes model to accommodate for certain properties of the order book in exchanges. 

\textbf{\textit{Prop 1}: Bid-Ask Spread is always non-negative:} \textcolor{black}{The bid-ask spread can never be less than one tick. This property is important to consider while modelling the LOB. This challenge was tackled by using order arrival intensities dependent on spread-in-ticks by \citealt{lee2022modeling}. Previously, \citealt{Zheng2014} have used constrained Hawkes Process to control for positive spreads. }

\textbf{\textit{Prop 2}: Order intensities are dependent on time-of-day:}  It is well known that trading volumes follow an intraday seasonality which is observered to be stationary across multiple days. Naturally, we observe that the order intensities too exhibit this intraday seasonality (more in Section 2). Hawkes models have been adapted to account for the same in \citealt{mucciante2023estimation}, \citealt{prenzel2022dynamic}, and \citealt{kirchner2022hawkes}. 

\textbf{\textit{Prop 3}: Cross-excitations can be inhibitory:} As noted in \citealt{LuAbergel2018}, the cross-excitation of events need not necessarily always be catalyzing. For example, we observe that limit orders inside the spread and cancels at top at opposing sides of the book have inhibitory effects on each other. Generally modelling for this can introduce negative intensities in the point process. To avoid such complications, \citealt{LuAbergel2018} propose to floor the total intensity of any element of the Hawkes process to zero. 

In this work we show some lack of support for the hypothesis that the order arrival intensities are impacted by the past order sizes. Thus we conclude that the Compound Hawkes Process is a suitable candidate for the model. We then create a stationary distribution of the order sizes for each type of order which closely mimics the empirical distribution. This distribution is used to sample the order sizes in the compound Hawkes Process. We create a novel formulation of the Hawkes intensities which satisfies Properties 1, 2 and 3. In Section 4, we show the calibration results for the Apple Inc. stock by using Level 2 data in the NASDAQ exchange. We also provide an analysis for the quality of fit as well as a market impact study in the simulator. 

\section{Related Work}

We refer to  \citealt{jain2024limit} for a comprehensive review of various techniques in Limit Order Book simulations. \textcolor{black}{\citealt{Gould2013} too reviewed Limit Order Books, focusing on their properties and presenting various models for LOB simulation. In another survey, \citealt{cont2011} demonstrated the usefulness of several zero-intelligence models in LOB modeling and provided empirical observations to test the models' outputs. For a more detailed overview of the microstructural statistics of the LOB and their modelling methods, we refer to the text by  \citealt{abergel_anane_chakraborti_jedidi_munitoke_2016}. The field of order book simulations is evolving alongside modeling techniques. With the rise of deep learning, many architectures have been developed to mimic the order book and its properties. \citealt{jain2024limit}'s review classifies order book simulator models based on their core methodologies:
\begin{enumerate}
    \item \textbf{Point Process Models}: These treat the order book as a queuing system with multiple components, which can be modeled either independently (zero-intelligence models) or with interacting terms (Hawkes process). They focus on aggregate dynamics, ignoring individual order motivations.
    \item \textbf{Agent-Based Models}: These simulate the behavior of various types of agents interacting with the central LOB. 
    \item \textbf{Deep Learning Models}: These avoid forming priors on data, using generative networks to estimate probability distributions. Although powerful, they face training difficulties and lack explainability.
    \item \textbf{Stochastic Partial Differential Equation Models}: Using observed facts or expert judgment, these models employ partial differential equations to describe LOB dynamics. 
\end{enumerate}
}%

The Hawkes Process has emerged as a promising model for addressing challenges associated with Poisson Processes in modeling Limit Order Book (LOB) queueing systems. In their comprehensive review, \citealt{Bacry2015} outline the key concepts of the Hawkes Process, its mathematical foundations, essential properties, and its applications, including an in-depth exploration of Order Book models. Notably, Hawkes Processes exhibit improvements over Poisson methods in capturing volatility clustering and addressing the Epps Effect, where covariance between assets approaches zero as the timescale decreases. The inherent features of endogenously excited order flow and implicit market impact make Hawkes Processes particularly relevant.

Recent work by \citealt{Hawkes2018} delves into the financial applications of Hawkes Processes, emphasizing their utility in modeling various market phenomena. Discussions in the literature have focused on selecting appropriate kernel functions, with studies demonstrating that power law kernels better fit empirical data than exponential ones (\citealt{nystrom2022hawkes}). \citealt{fonseca2014calib} argue against exponentially decaying kernels based on Q-Q plot comparisons, highlighting potential insufficiencies in representing empirical data.

The extension to n-dimensional Hawkes Processes has gained popularity in LOB models, with different formulations. \citealt{toke2010market} present a two-agent model employing 1D Hawkes Processes for Market Orders and Limit Orders.  \citealt{bacry2016estimation} divide order book events into categories, using an 8-dimensional Hawkes Process to model bid and ask sides separately.  \citealt{kirchner2017estimation} propose a non-parametric method for estimating Hawkes Processes, optimizing hyperparameters using the AIC statistic.  \citealt{fonseca2014calib} offer an alternate strategy for fitting Hawkes Processes, utilizing a generalized method of moments for faster parameter estimation compared to traditional maximum likelihood methods. They validate their approach using key stylized facts and compare results with MLE baselines.

In the realm of Constrained Hawkes Processes,  \citealt{Zheng2014} and  \citealt{lee2022modeling} present 4-dimensional models with constraints on bid-ask spread processes. \textcolor{black}{\citealt{MuniToke2017, muniToke2020} also present a spread dependent LOB point process model for European (Large-tick) stocks.} \citealt{lee2022modeling} introduce stochastic decay kernels and provide empirical validation and economic interpretations. Various other variants of Hawkes Processes have been proposed. \citealt{kaj2017buffer} introduce a "Buffer-Hawkes" Process, while  \citealt{pakannen2022} develop a state-dependent Hawkes Process based on spread and order flow imbalance. \citealt{kirchner2022hawkes} formulate a marked state-dependent Hawkes Process, optimizing for model parsimony.

\textcolor{black}{The LOB's individual order sizes have been a quantity of interest in a number of other works. \citealt{Gopikrishnan2000} observe a power law decay in the probability distribution of traded volume for several equities, which is the order size of Market Orders in our model.  With regards to order size modelling within an LOB model, let us mention that \citealt{abergel_anane_chakraborti_jedidi_munitoke_2016} use parametric distributions for order sizes in the Hawkes Process model. The fact that order sizes distributions are not smooth and have spikes in probability at round numbers, probably due to the traders' preference for round numbered child order sizes, has been noted in \citealt{CHALLET2001285}. More recently, \citealt{lu2018order} develop geometric distributions with spikes at round numbers for their Poisson LOB model's order sizes. In an alternate direction, a non-parametric method for sampling order sizes was used in \citealt{Hultin2023} where each order size is a result from a deep recurrent neural network.}

Scaling limits of Hawkes Processes have been explored by  \citealt{Abergel2015} and \citealt{horst2019scaling}, demonstrating convergence to Stochastic Differential Equations (SDEs) and Ordinary Differential Equations (ODEs) for aggregate features. Non-linear Hawkes Processes, as proposed by  \citealt{lu2018order} and \citealt{mounjid2019asymptotic}, introduce inhibiting kernels for negative excitation and novel features such as sum of exponential functions. These models offer enhanced performance compared to linear counterparts. The application of Neural Hawkes Processes has gained attention, with \citealt{kumar2021deep} introducing a Deep Neural Hawkes Process for Market Making. \citealt{shi2022state} develop a neural Hawkes process with continuous time LSTM units, capturing complex dynamics in market feedback loops.

Hawkes Processes offer a comprehensive point process methodology for modeling LOB arrivals, showcasing adaptability and the ability to capture microstructural details. Despite their strengths however, challenges exist in model calibration, especially due to the complex likelihood function and the impact of kernel choice on predictive power. Model parsimony also remains a pertinent consideration in higher-dimensional settings.

\section{Methodology}
\subsection{Mathematical Formulation:}  
We consider the task of modelling LOB's current best bid and ask (i.e. Level 1 LOB). We note that since placing order in the spread is a common practice in equities, the reason being that the spread usually has empty price levels \textcolor{black}{(the number of empty price levels in the spread varies for each equity, particularly notable is the finding by \citealt{bouchaud2018trades} where the authors show a strong relationship between the price of US stocks and the spread-in-ticks)} to attract market participants to place a more aggressive order, the Hawkes formulation should contain modelling the intensity of the nearby empty levels as well. Hence we formulate the order book as a queueing system with six different queues which are $\{\text{ask}_{+1}, \text{ask}_0, \text{ask}_{-1}, \text{bid}_{+1}, \text{bid}_0, \text{bid}_{-1}\}$. 

\begin{figure}
       \centering
       \includegraphics[width=0.4\textwidth]{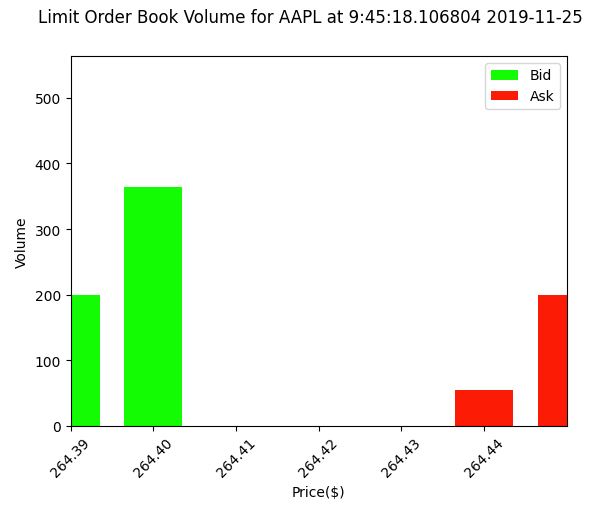}
       \caption{Order Book Snapshot}
       \label{fig:lobSnapshot}
   \end{figure}
 
Here $\text{ask}_0$ and $\text{bid}_0$ are the respective best ask and bid prices while the subscripts denote the price level distance in ticks from the best ask/bid. For eg. $\text{ask}_{+1}$ is the price level which is 1 tick more than $\text{ask}_0$. The $2 \times 6$ vector of $((q^{(\zeta)}_{k} , \zeta_k)_{\zeta \in \{\text{ask}, \text{bid}\}, k \in \{-1, 0, +1\}})$, where $q^{(\zeta)}_k$ denotes the queue size, is what we define as the state of the order book. The motivation to model just these six levels and not the entire order book is two-fold. Firstly, the question of parsimony becomes important when we model more levels than just these six. Secondly, we observe that the LOB state changes are generally events which move the prices by 1 tick. Indeed we observe, in our dataset, 97.4\% of all in-spread orders occur 1-tick away from the best respective quote with mean being $1.035 \pm 0.285$ ticks for 2 million samples, and 97.5\% of all price changes have the next non-empty price level at 1-tick distance as well with the mean being $1.033 \pm 0.283$ ticks. 

\textbf{Mechanical Constraints:} Naturally, since $\text{ask}_0$ and $\text{bid}_0$ are the best ask and bid prices, the queue size at $\text{ask}_{-1}/\text{bid}_{+1}$ is zero. Therefore any incoming limit order at $\text{ask}_{-1}/\text{bid}_{+1}$ (i.e. an in-spread (IS) order) creates a new best ask/bid and so the LOB state transforms from 

\begin{align*}
     S : = \{ &(q^{(\text{ask})}_{+1}, \text{ask}_{+1}), (q^{(\text{ask})}_0,\text{ask}_0), (0,\text{ask}_{-1}), \\ & (0, \text{bid}_{+1}), (q^{(\text{bid})}_{0}, \text{bid}_0),(q^{(\text{bid})}_{-1}, \text{bid}_{-1})\}
\end{align*}
to \begin{align*}
    S_{IS} = \{& (q^{(\text{ask})}_0,\text{ask}_0), (q^{(\text{ask})}_{-1},\text{ask}_{-1}), (0, \text{ask}_{-2}), \\ & (0, \text{bid}_{+1}), (q^{(\text{bid})}_{0}, \text{bid}_0),(q^{(\text{bid})}_{-1}, \text{bid}_{-1})\}
\end{align*} for an in-spread ask side limit order.

 Note that it is possible that the spread is only \textcolor{black}{2-tick} wide which would mean $\text{ask}_{-1}$ and $\text{bid}_{+1}$ coincides. There is also the possibility of \textcolor{black}{1-tick} spread in which case $\text{ask}_{-1}$ and $\text{bid}_{+1}$ do not exist. We will put some constraints on the Hawkes process intensities for these levels further in this section to account for these possibilities. On the other hand, if a queue-depletion (QD) event at best bid/ask happens (for eg, a large enough market order of size $\kappa^{MO_\text{ask}} \geq q^{(\text{ask})}_0$), the spread widens and therefore the state moves from $S$ to  \begin{align*}
     S_{QD} = \{&(q^{(\text{ask})}_{+2} - (q^{(\text{ask})}_0 - \kappa^{MO_\text{ask}}), \text{ask}_{+2}), (q^{(\text{ask})}_{+1},\text{ask}_{+1}), (0,\text{ask}_{0}), \\&(0, \text{bid}_{+1}), (q^{(\text{bid})}_{0}, \text{bid}_0),(q^{(\text{bid})}_{-1}, \text{bid}_{-1})\}
 \end{align*} for a queue-depleting ask side market order/cancel order. Here we observe an unknown quantity $q^{(\text{ask})}_{+2}$. We sample this unknown quantity from a stationary distribution calibrated from empirical data. \textcolor{black}{We remove the remainder of the Market Order's size (if any) from this quantity to get the final state $S_{QD}$. We stress the fact that Market Order sizes relative to the quoted liquidity can be high enough to incur a multi-price-level QD, in which case the spread widens by multiple ticks and the state moves accordingly.} Finally, we note that Market Orders can only occur at $\text{ask}_0 / \text{bid}_0$ and $\text{ask}_{-1} / \text{bid}_{+1}$ cannot have Cancel Orders since by definition the quantity there is zero.

\textbf{Compound Hawkes Process:} A compound point process (CPP) $Z(t)$ is defined as 
\textcolor{black}{
\begin{align}
Z(t) = \sum^{N(t)}_{i = 1} Y_i  
\end{align} 
}%
where $\{N(t) : t \geq 0\}$ is the counting process, \textcolor{black}{with intensity $\lambda(t)$,  and $Y_i$ are i.i.d. random variables denoting the jump size with the corresponding $P_{\Theta}(Y_i)$ being their respective jump size distribution parameterised by $\Theta$. } Compound Hawkes process (CHP) is a special case of the CPP where $N(t)$ is a counting process associated with a Hawkes Process. For a $d$-dimensional Hawkes process the intensity of the process \textcolor{black}{$\lambda^{(i)}(t)$ and the associated counting process $N^{(i)}(t)$ for $ i = 1,\ldots, d$ is defined as: 
\begin{align}
    \lambda^{(i)}(t)  = \mu^{(i)}(t) + \sum^d_{j=1} \sum_{t_j \in T_j} \phi^{(j \rightarrow i)}(t - t_j)
\end{align}} where $T_j := \{t_j : t_j \leq t\}$ denotes the set of past event times in the $j$ dimension of the Hawkes Process. Here, \textcolor{black}{$\mu^{(i)}(t)$} is the exogenous intensity of the $i$-th dimension and $\phi^{(j \rightarrow i)}(t - t_j)$ is the excitation term from $j$-th dimension to $i$-th dimension. The excitation terms are a function of the time since the event (generally a decaying function in time like exponential decay or power law decay). An alternate but similar formulation is the following: 
\textcolor{black}{
\begin{align}
    \lambda^{(i)}(t)  = \mu^{(i)}(t) + \sum^d_{j=1} \int^t_{0} \phi^{(j \rightarrow i)}(t - u) dN^{(j)}(u)
\end{align}
}%
\textbf{Limit Order Book at Compound Hawkes Process:} We define a 12D Compound Hawkes Process (in accordance with the mechanical constraints) for the 6 queues in the order book state where each dimension corresponds to the following event types:  

\begin{align*}
\mathcal{E} := \{LO_{\text{ask}_{+1}}, CO_{\text{ask}_{+1}}, LO_{\text{ask}_{0}}, CO_{\text{ask}_{0}}, MO_{\text{ask}_{0}}, LO_{\text{ask}_{-1}}, \\ 
LO_{\text{bid}_{+1}}, LO_{\text{bid}_{0}}, CO_{\text{bid}_{0}}, MO_{\text{bid}_{0}}, LO_{\text{bid}_{-1}}, CO_{\text{bid}_{-1}}\}
\end{align*}
where LO is Limit Order, CO is Cancel Order and MO is Market Order. \textcolor{black}{We define a one-to-one mapping between the enumeration $i = 1, \ldots, 12$ to the ordered set $\mathcal{E}$ denoted by $i(e):\mathcal{E} \rightarrow \{1, \ldots, 12\}$.} Here each of the 12 event types' order size (\textcolor{black}{$\kappa^{(e)} \in \mathbb{N}$} for event \textcolor{black}{ $e\in \mathcal{E}$}) is sampled from its own specific calibrated distribution ($\Pi^{(e)}(\Theta)$ where $\Theta$ is its calibrated parameters). We postulate that the order intensities themselves are not impacted by the past order sizes but only the past order event-times as is the general assumption in Hawkes models applied to LOB data. We provide some weak evidence for this claim in the Appendix A. We note that our work on the distribution of order size closely follows the work by \citealt{lu2018order} however they make use of a Poisson model. For Cancel Order quantities, since we observe almost all cancels in empirical data are outright cancels, we draw randomly, with equal probability, from the available limit orders' quantities in the queue. We thus create a CHP for the queue-sizes at the six queues\textcolor{black}{, indexed by $k = 1 \ldots 6$,} of the LOB:

\textcolor{black}{
\begin{align}
    q^{(\zeta)}_k(t) &= \sum_{e \in \mathcal{E}^{(\zeta)}_k}\sum^{N^{(i(e))}(t)}_{j=1} \kappa^{(e)}_j\\
    \kappa^{(e)}_j &\sim \Pi^{(e)}(\Theta) \\  &\forall j \in \mathbb{N}, \zeta \in \{\text{ask, bid}\}, e \in \mathcal{E}, \kappa^{(e)}_i \in \mathbb{N} \nonumber
\end{align}
}%
where $j = 1, \ldots, N^{(i(e))}(t)$ and $\mathcal{E}^{(\zeta)}_k$ is the set of eligible events for \textcolor{black}{side $\zeta$ and queue number $k$}.

\textcolor{black}{The assumption that these distributions are stationary and independent of any conditions is quite strong. Addition of queue-reactiveness is a potential solution to this problem. We reserve this line of research for future work.} 

To control the spread of the process (Prop 1), we formulate the intensities of $LO_{\text{ask}_{-1}}$ and $ LO_{\text{bid}_{+1}}$ in the following manner as a function of the current spread-in-ticks ($s$). \textcolor{black}{If $\lambda^{(IS)}(t)$ represents the intensity of the In-Spread events : $\{i(LO_{\text{ask}_{-1}}), i(LO_{\text{bid}_{+1}})\}$, then, \begin{align}
   \lambda^{(IS)}(t,s) =  \lambda^{(IS)}(t)  (s-1)^\beta;  \beta > 0, s \in \mathbb{N} 
\end{align} 
}%
In this formulation we enforce the mechanical constraint if \textcolor{black}{$s = 1$, $\lambda^{i(LO_{\text{ask}_{-1}})}(t, 1) = \lambda^{i(LO_{\text{bid}_{+1}})}(t, 1) = 0$} with an extra parameter $\beta$. We motivate the choice of a power law dependence of order intensity over spread in Appendix B.

We observe a strong dependence between the intensities $\lambda^{(.)}_t$ on the time of day of the trading day (Prop 2). There is a "U"-shape of order intensities with respect to the time of day i.e. right after the open auction and before the close auction, the activity in the market is much higher than the middle of the day. This effect is observed for trading volumes in equities with a number of probable causes but the most commonly accepted one is that because auctions cause a halt in trading, this halt leads to increased activity. To maintain model parsimony, we bin the 6.5 hour trading day of NASDAQ into thirteen 30-minute bins (indexed by $Q_t \in \{1,2,..., 13\}$) for $t$ being a time of day). We model the intensity as a separable function of time-of-day and the Hawkes Intensities: 

\begin{align}
    \lambda^{(i)}(Q_t, t, s) = f^{(i)}(Q_t) \times \lambda^{(i)}(t, s)
\end{align}

Finally, to make sure the inhibitory kernels do not create negative intensities, we floor the total excitation $\lambda^{(e)}(Q_t, t, s)$ to zero. Therefore the generalized intensities for all 12 dimensions, $i = 1 \ldots 12$, can be written as:

\textcolor{black}{
\begin{align}
\text{For In-Spread events:} & \nonumber \\
    \lambda^{(i)}(Q_t, t, s)  =& 	max \bigg(0, f^{(i)}(Q_t) (s-1)^\beta \times \nonumber \\& \bigg( \mu^{(i)} + \sum^d_{j=1} \int^t_{0} \phi^{(j \rightarrow i)}(t - u) dN^{(j)}(u) \bigg) 	\bigg) \\
 \text{For other event :} \qquad & \nonumber \\
    \lambda^{(i)}(Q_t, t, s)  =& max \bigg(0,f^{(i)}(Q_t) \times \nonumber \\&\bigg( \mu^{(i)}  + \sum^d_{j=1} \int^t_{0} \phi^{(j \rightarrow i)}(t - u) dN^{(j)}(u) \bigg) \bigg)
\end{align}
}%

\textbf{Price Dynamics:} The mid price ($P(t)$) dynamics after an event at $t$ at side $\zeta$ (and $\xi$ = +1 for $\zeta = $ ask and -1 for $\zeta = $ bid) is a jump process with jumps defined by :

\begin{align}
P\left(t^{+}\right) - P(t) = & -\frac{\eta}{2} \xi d N^{(I S)}(t) \nonumber\\ & + \frac{\eta}{2} \xi d N^{(C O)}(t) \mathbb{1}_{\{\# q_0^{(\zeta)}(t)=1\}} (t) \nonumber\\ & + 
\frac{\eta}{2} \xi d N^{(M O)}(t) \mathbb{1}_{\{q^{(\zeta)}_0(t) \leq \kappa^{(M O_{\zeta_0})}(t)\} }(t)
\end{align}
where $\eta$ is the tick-size, $N^{(I S)}(t)$ is the counting process for IS orders, and $\# q_0^{(\zeta)}(t)$ denotes the number of individual limit orders in the queue. 

\subsection{Calibration:}

We follow the non-parametric method of calibration in \citealt{kirchner2017estimation} and take inspiration from \citealt{bacry2016estimation} to create a time grid at both linear and log scales to account for the slowness of the decay kernels. This method has the advantage of not having any priors on the shape of the kernels themselves (popular choices in the literature include exponential decay and power-law decay) since it is non-parametric. 

{For an event $e$ (not a spread-dependent event), with the counting process $N^{(i(e))}(t)$ having an intensity as in (9), we define the binned observated counts in the dataset by $X^{(e, n)}_{\Delta} := N^{(i(e))}(t_n) - N^{(i(e))}(t_{n-1})$, $n \in \mathbb{N}$ where $\Delta := t_n - t_{n-1}$ is the bin size (taken constant). For brevity, we drop the notation of $i(e)$ and instead denote the index by the event $e$ itself. Taking inspiration from \citealt{kirchner2017estimation} we have, for $\mathcal{H}^{(e)}_{n-1}$ denoting the set of previous $n-1$ bins' observed values:
{\small
\begin{align}
    \mathbb{E}\bigg[ X^{(e, n)}_{\Delta} \bigg| \mathcal{H}^{(e)}_{n-1} \bigg] =& \int^{n\Delta}_{(n-1)\Delta}  \mathbb{E}\bigg[ f(Q_t) \lambda^{(e)}(t) \bigg| \mathcal{H}^{(e)}_{n-1} \bigg] \\
    \frac{\mathbb{E}\bigg[ X^{(e,n)}_{\Delta} \bigg| \mathcal{H}^{(e)}_{n-1} \bigg]}{f(Q_t)}  =& \int^{n\Delta}_{(n-1)\Delta} \mu^{(e)}_t dt  \nonumber \\ 
    &+ \sum^{d}_{j = 1} \int^{n\Delta}_{(n-1)\Delta} \int^{t}_{0} \phi^{(j\rightarrow i)}(t-u) N^{(j)}(du) dt 
\end{align}
}%
}%
{Here we use the fact that $f(Q_t)$ is piecewise constant and construct the bins such that there is \textcolor{black}{perfect} overlapping between the $\Delta$ grid and the $Q_t$ grid. Now we define another grid $\delta_k :=  t^{(k)}_{max} - t^{(k)}_{min}$ with \textcolor{black}{ $t^{(1)}_{max} = t; t^{(K)}_{min} = t - T$ } for $ k = 1, ..., K$ and $T < \infty$ being the cut-off point for the integral to make an approximation:
{\small
\begin{align}
\int^{t}_{0} \phi^{(j\rightarrow i)}(t-u) N^{(j)}(du) &\approx \sum^{K}_{k = 1} \frac{X^{(e)}_{\delta_k}}{\delta_k} \int^{t^{(k)}_{max}}_{t^{(k)}_{min}} \phi^{(j\rightarrow i)}(u) du
\end{align}
}%
}%
{Here $X^{(e)}_{\delta_k}$ is the binned observed events of type $e$ in the bin $\delta_k$. Now we make another approximation, as in \citealt{kirchner2017estimation}, that the grid $\Delta$ is small enough so that the intensity is constant over $(t_{n-1}, t_n ]$. Now with this and substituting (11) in (10), we have:
{\small
\begin{align}
\frac{\mathbb{E}\bigg[ X^{(e, n)}_{\Delta} \bigg| \mathcal{H}^{(e)}_{n-1} \bigg]}{f(Q_t)}  \approx& \Delta \mu^{(i)} + \Delta \sum^d_{j = 1} \sum^{K}_{k = 1}\frac{X^{(e)}_{\delta_k}}{\delta_k} \int^{t^{(k)}_{max}}_{t^{(k)}_{min}} \phi^{(j\rightarrow i)}(u) du
\end{align}
}%
}%
{As noted in \citealt{bacry2016estimation}, a linearly spaced grid to estimate cross-excitation is insufficient for slow decaying kernels like the power-law. Therefore we propose to use the same linear-log spaced grid proposed in \citealt{bacry2016estimation} as the $\delta_k$ grid. We set $\delta_1 = \Delta$ and add $\frac{K}{2}$ linearly spaced $\delta_k$. Lastly we add $\frac{K}{2}$ log spaced $\delta_k$ to get the final grid $G = [0, \Delta, ..., \frac{K}{2}\Delta, \frac{K}{2}\Delta m,  \frac{K}{2}\Delta m^2,...,T]$ with $m = (\frac{2T}{K\Delta})^{\frac{2}{K}}$. Note that for in-spread events, the intensity also depends on the current spread. However we note that much like $f(Q_t)$, \textcolor{black}{$(s_t - 1)^\beta$} is another piecewise constant function multiplied to the intensity. Therefore the treatment of in-spread events is exactly the same with $f(Q_t)$ being replaced by \textcolor{black}{$f(Q_t) \times \mathbb{E}[(s_t - 1)^\beta]$}. For in-spread events, we have, for current spread $s$ in the $\Delta$ bin event $e$, the following equations. Here \textcolor{black}{$ <(s-1)^\beta>_{((n-1)\Delta, n \Delta]} $} denotes the average of \textcolor{black}{$(s-1)^\beta$} in the bin $\Delta$. The same procedure as above follows for the modified AR process fitting of the in-spread events' intensities.
\textcolor{black}{
\begin{align}
    \mathbb{E}\bigg[ X^{(e, n)}_{\Delta} \bigg| \mathcal{H}^{(e)}_{n-1} \bigg] = \int^{n\Delta}_{(n-1)\Delta}  &\mathbb{E}\bigg[ (s-1)^\beta f^{e}(Q_t) \lambda^{(e)}(t) \bigg| \mathcal{H}^{(e, N)}_{n-1} \bigg]
\end{align}
\begin{align}
    \frac{\mathbb{E}\bigg[ X^{(e,n)}_{\Delta} \bigg| \mathcal{H}^{(e)}_{n-1} \bigg]}{f^{(e)}(Q_t)<(s-1)^\beta>_{((n-1)\Delta, n \Delta]}}  =& \int^{n\Delta}_{(n-1)\Delta} \mu^{(e)}_t dt  \nonumber \\ 
    + \sum^{d}_{j = 1} \int^{n\Delta}_{(n-1)\Delta} & \int^{t}_{0} \phi^{(j\rightarrow i)}(t-u) N^{(j)}(du) dt
\end{align}
}%
This formulation is then solved by the least squares method. We get as a result (upon rescaling by $\frac{\delta_k}{\Delta}$) the point estimates for \\$\int^{t^{(k)}_{max}}_{t^{(k)}_{min}} \phi^{(j\rightarrow i)}(u) du$ for all $t^{(k)}_{(.)} \in G; i, j \in {1, ..., d}$. We form cumulative sums of these estimates to get point estimates for the function $\Phi(t^{(k)}_{max}) = \int^{t^{(k)}_{max}}_{0} \phi^{(j\rightarrow i)}(u) du$. 

We add two constraints to the least squares problem (from the auto-regressive formulation). Firstly, we enforce the unidirectional shape of the Hawkes Kernels by constraining the point estimates to be uniform in sign (i.e., either all greater than 0 or all less than 0). For this, we make use of the Hawkes Graph \citealt{achab2018uncovering} which directly estimates the kernel norms matrix : $\boldsymbol{M}_{d \times d}$ (elements of which are defined as $M_{ij} = \int_0^\infty \phi^{(j \rightarrow i)}(u)du$ which is the kernel norm of $\phi^{(j \rightarrow i)}(t)$ for $i, j \in \{1, ..., d\}$) from the conditional probabilities of the 12D point process. Note that we purely use this method to identify whether a kernel is catalysing or inhibiting in nature (i.e. is the kernel norm positive or negative). 

Secondly, we constrain the integral of a Hawkes Kernel to ensure the stability of the Hawkes Process. A standard Hawkes Process is stable if the maximum eigenvalue of the Kernel Norm matrix is $<1$ (\citealt{Hawkes2018}), we derive the corresponding constraint for our Hawkes Process model in Equations 17-21. 
\begin{align}
     \lambda^{(i)} & = \mathbb{E}\bigg[\lambda^{(i)}(t)\bigg]\nonumber \\ 
     & = \mathbb{E}\bigg[f^{(i)}(Q_t) \bigg(\mu^{(i)}_t + \sum^d_{j=1} \int^t_{0} \phi^{(j \rightarrow i)}(t - u) dN^{(j)}_u\bigg)\bigg]   \\
     & = f^{(i)}(Q_t) \bigg(\mu^{(i)} + \sum^d_{j=1} \lambda^{(j)}\int^t_{0} \phi^{(j \rightarrow i)}(t - u) du\bigg)\\
     \Lambda & = \text{diag}(f^{(1)}(Q_t),\ldots,f^{(d)}(Q_t)) \cdot \bigg(\boldsymbol{\mu} + \Lambda \cdot \boldsymbol{M}\bigg)\\
      \Lambda & = \boldsymbol{\mu}\bigg(\text{diag}(f^{(1)}(Q_t),\ldots,f^{(d)}(Q_t))^{-1} - \boldsymbol{M}\bigg)^{-1}\\
    \implies   \int^{\infty}_{0} &\phi^{(j\rightarrow i)}(u) du  \approx \sum^{K}_{k = 1}  \int^{t^{(k)}_{max}}_{t^{(k)}_{min}} \phi^{(j\rightarrow i)}(u) du   \leq \frac{1}{max_{Q_t} f(Q_t)}
\end{align}
Therefore for stationarity and stability of the Hawkes Process, the maximum eigenvalue of the matrix $\boldsymbol{M}$ should be less than $\frac{1}{max_{Q_t}f(Q_t)}$. We postulate that the formulation with constrained integrals of individual kernels (Equation 19) is an adequate linear approximation to the non-linear constraint of bounded eigenvalues. In practice, we do see the calibrated Hawkes Kernel to occasionally violate this non-linear constraint. We apply a heuristic of multiplying all the kernels by a regularizing term to cap the eigenvalue at $1$. 

Following the same logic as equations 13 and 14, one can get the following constraint for in-spread order intensities.
\textcolor{black}{
\begin{align}
    \sum^{K}_{k = 1}  \int^{t^{(k)}_{max}}_{t^{(k)}_{min}} \phi^{(j\rightarrow i)}(u) du   & \leq \frac{1}{ < (s-1)^\beta> max_{Q_t} f(Q_t)}
\end{align}
}%
These constraints are linear constraints on the Quadratic Problem posed by equation 12. 

We fit parametric functions on each of these kernels on these point estimates and here we make use of power-law functions and exponential functions as the candidates. Unlike \citealt{kirchner2017estimation}, however, we do not calibrate the parameters in 30-minute windows and average over a day's 13 windows, but rather we use the entire day's data to calibrate the parameters and secondly we do not make the approximation $\int^{t^{(k)}_{max}}_{t^{(k)}_{min}} \phi^{(j\rightarrow i)}(u) du \approx (t^{(k)}_{max}  - t^{(k)}_{min} ) \phi^{(j\rightarrow i)}(t^{(k)}_{min}) $ since the grid size is not uniformly small. 
 
We calibrate the data for several days individually and test the stationarity (by day) of the calibrated parameters. Since we do see stationarity in the kernel parameters over multiple days (Figure \ref{fig:kernelShape}), we conclude by using the parameters over multiple days for our final calibrated parameters' estimation. 
We note that as mentioned in \citealt{kirchner2018mle}, the exogenous intensity is massively overestimated because of the cut-off error and the distributional errors. Therefore me make use of the result from \citealt{bacry2016estimation} to estimate the exogenous intensities from the average intensity of the events observed ($\hat{\Lambda}$) and the fitted norm matrix $\hat{M}$. From Equation 17, we take the average across all bin indices $Q_t$ and use the fact that $<f(Q_t)> = 1$ to get:
\begin{align}
\hat{\boldsymbol{\mu}} & = ( \mathbb{I}- \hat{\boldsymbol{M}})\hat{\Lambda}
\end{align}
}%
We use the maximum likelihood method to calibrate the order size distribution. A key observation is the preference of traders of round numbers in their order quantity. This is an important stylized fact of the order book dynamics so we make use of Dirac delta functions to add spikes in the distribution function to account for this. Finally, we use the thinning algorithm to simulate the order book from the calibrated parameters and provide some visualizations of the calibrated parameters in the following section along with the quality of fit results.


\section{Results}

\textcolor{black}{In this section we showcase the calibrated and simulated results of our proposed model. We further perform comparisons to other models of LOB in the literature. For a comparison of the non-parametric method of calibration of the Hawkes Process against the more commonly used MLE method, we refer the reader to the work by \citealt{kirchner2018mle}.}

\subsection{{Calibration for Simulated Data:}}

{As evidence of correctness of the fitting method, we construct a 12D Compound Hawkes Process and simulated it for 100 trading days. The parameters of this CHP are chosen randomly from a list of candidate parameters and the norm matrix is chosen to be diagonal with near-diagonal terms having non-zero elements. The real norm matrix is shown in Figure \ref{fig:realM_fakeData}. We then transform the simulated order book data into the binned data structure needed for the calibration methoodology above and perform the non-parameteric estimation of the kernel point estimates and the exogenous intensities, and finally fit a power law kernel to the points observed over the 100 trading days. We show the fitted norm matrix in Figure \ref{fig:fittedM_FakeData}.
\begin{figure}
\begin{subfigure}{.5\columnwidth}
  \centering
  \includegraphics[width=.9\linewidth]{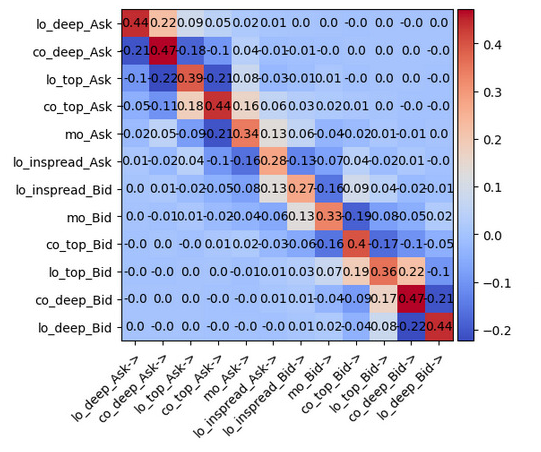}
  \caption{Norm Matrix: Real, Simulated Data}
  \label{fig:realM_fakeData}
\end{subfigure}%
\begin{subfigure}{.5\columnwidth}
  \centering
  \includegraphics[width=.9\linewidth]{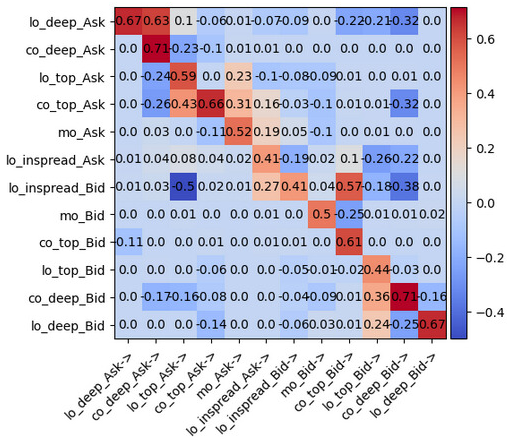}
  \caption{Norm Matrix: Fitted, Simulated Data}
  \label{fig:fittedM_FakeData}
\end{subfigure}
\caption{Simulated Data Calibration}
\label{fig:realfittedM_FakeData}
\end{figure}
We note that the fitted matrix exhibits similar shape near the diagonal to the real matrix however we do see some off diagonal elements having large inaccurate values. The norm weighted average relative error in estimated parameters across the 144 kernels is 6.0\%, 0.9\% and 4.4\% respectively for $\alpha, \beta, \gamma$ parameters of the power law kernel $\phi(t) = \alpha (1 + \gamma t)^{-\beta}$ . Hence we conclude that the fitting methodology works reasonably well with a tolerable amount of estimation error.
}%
\subsection{Calibration for Equities LOB:}

\begin{figure}
\centering
\includegraphics[width=0.5\linewidth]{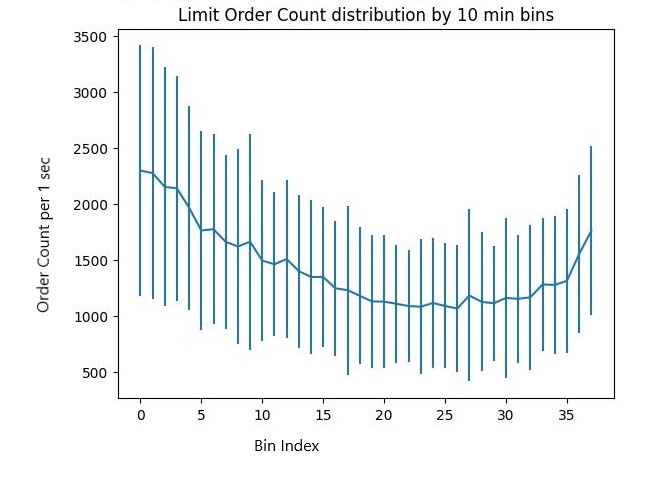}
\hfill
\caption{Intensities conditioned by Time of Day : we report the average number of events per 30 min bin in a 6 month period}
\label{fig:exoTOD}
\end{figure}

\begin{figure}
    \centering
    \begin{subfigure}{0.49\textwidth}
    \includegraphics[width=\linewidth]{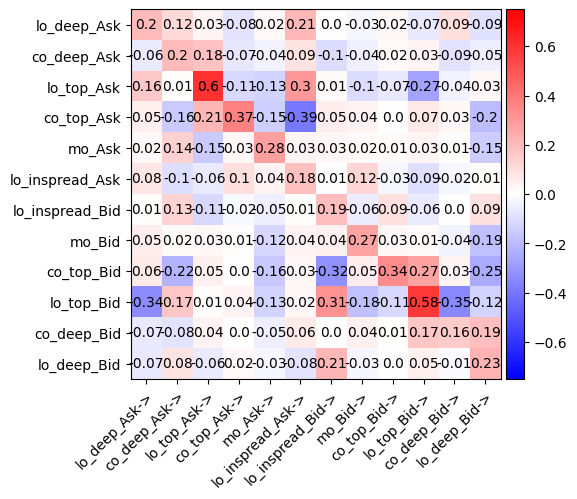}
    \caption{Norms of Kernel: AAPL.OQ}
    \label{fig:kernelNorms}
    \end{subfigure}
    \hfill
    \begin{subfigure}{0.49\textwidth}
    \centering
    \includegraphics[width=\linewidth]{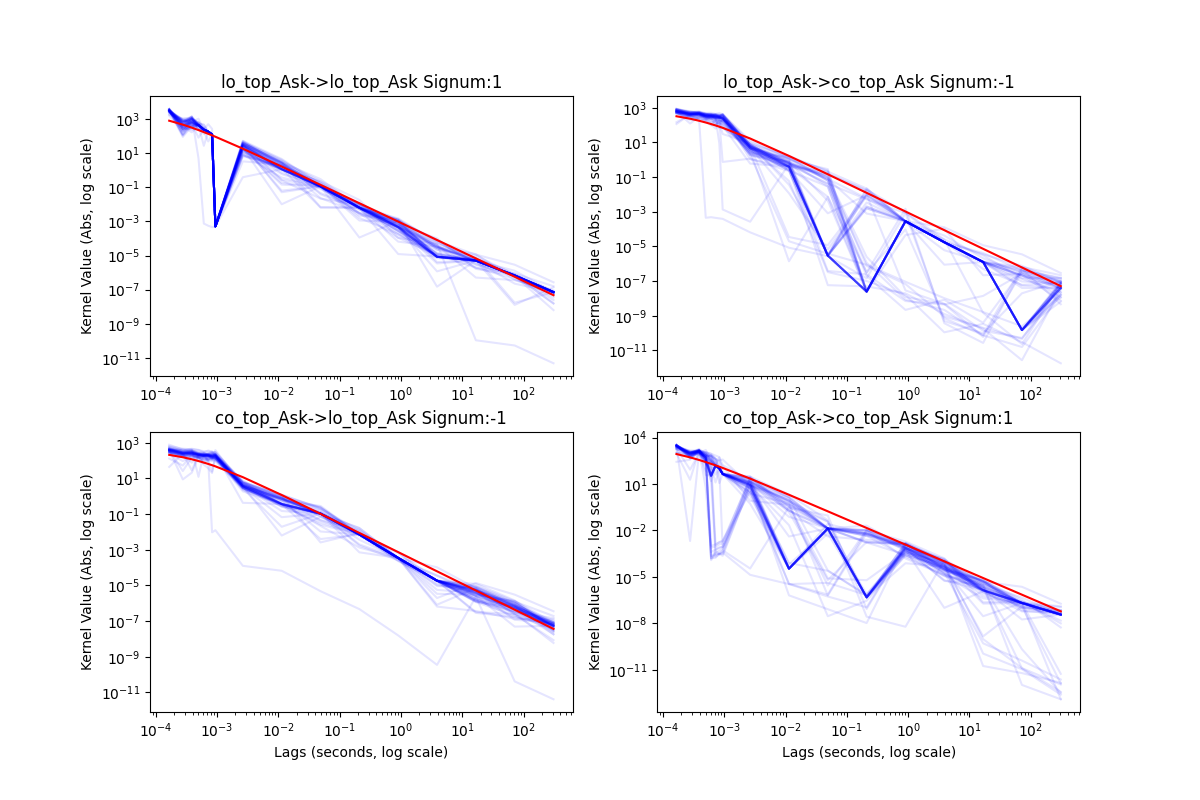}
    \caption{Kernel Shape: non-parametric point estimates (blue) and parametric power-law fit (red) - (log-log scale) : AAPL.OQ}
    \label{fig:kernelShape}
    \end{subfigure}
    \caption{Excitation Kernels}
\end{figure}

We now show the observed intensities conditioned by time of day for a sample dimension (Market Orders at Ask) in Figure \ref{fig:exoTOD}. We observe the common U-shape across all 12 dimensions. In Figure \ref{fig:kernelNorms}, we show the the norm of kernels $\boldsymbol{M}$, removing the time-of-day multiplier. The $x$ axis shows the excitor and $y$ axis is the excitee. We also show some sample fitted kernel shapes in Figure \ref{fig:kernelShape}. As we can see here in the translucent blue lines, the kernel shapes fitted over multiple days are stable across days. 

Since the number of excitation kernels in a 12D Hawkes Process are 12$^2$, in order to maintain model parsimony, we zero-out the effect of small cross-excitations such as Cancel Bid$_0$ $\rightarrow$ Limit Ask$_0$. We set the threshold to select kernels on the basis of their norms to 0.01 as it eliminates several cross-excitation terms to limit the number of kernels to 65. 

Note that these estimated kernels are estimated in a non-parametric manner and hence need to be further fitted on a parametric function. We choose between the power-law kernel : $\phi(t) = \alpha (1 + \gamma t)^{-\beta}$ and the exponential kernel : $\phi(t) = \alpha e^{-\beta t}$ to do the parametric fit by comparing the fit's Akaike Information Criteria (AIC). We see that the power-law kernel is selected 100\% of the time for this dataset. We show the fitted line in \ref{fig:kernelShape} in red. As we can see, the power law line fits the point estimates quite well. Indeed we see an average mean square fit error to be $\sim 10^{-3}$ . Another noteworthy aspect of the fitting results is the presence of inhibitory kernels. 

We make use of NASDAQ Level 2 data for the year 2019 (i.e. 12 months of data) for Apple Inc. which is a medium-tick stock (average spread is 1.7 ticks). In Appendix C, we also show some more results for other stock types like large-tick (Intel), small-tick (Tesla) and very small-tick (Amazon). 


\subsection{Calibration of Order Size Distributions}

\begin{figure}
\begin{subfigure}{.5\columnwidth}
  \centering
  \includegraphics[width=.8\linewidth]{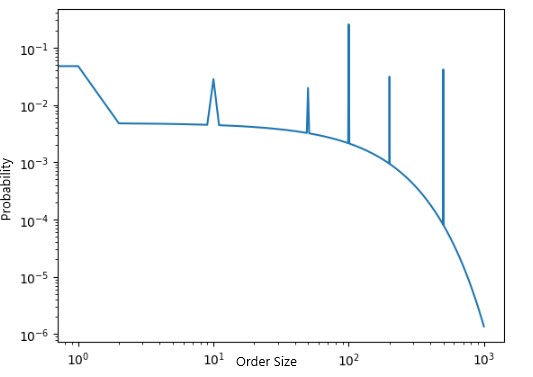}
  \caption{Market Orders' Size Distribution (log-log scale)}
  \label{fig:sfig1}
\end{subfigure}%
\begin{subfigure}{.5\columnwidth}
  \centering
  \includegraphics[width=.8\linewidth]{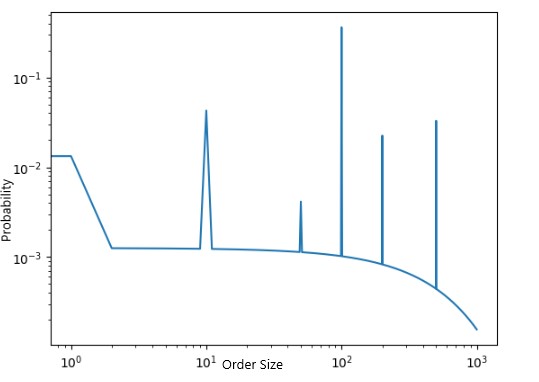}
  \caption{Limit Orders' Size Distribution (log-log scale)}
  \label{fig:sfig2}
\end{subfigure}
\caption{Fitted Distribution of Order Sizes}
\label{fig:fittedSizePDF}
\end{figure}

Following \citealt{lu2018order}, we use Dirac delta at round numbers to account for the stylized facts we observe in Figure \ref{fig:empMO} and \ref{fig:empLO}. We choose the set {1, 10, 50, 100, 200, 500} as the set of round numbers we wish to put spikes in the PDF at. The remainder of the PDF is modelled by a Geometric distribution. We fit this distribution using the maximum likelihood method. We show the final calibrated PDF (smoothened for illustration purposes) in Figure \ref{fig:fittedSizePDF}.


\subsection{Quality of fit metrics:}

\begin{figure}
\begin{subfigure}[c]{.32\linewidth}
  \centering
  \includegraphics[width=.95\linewidth]{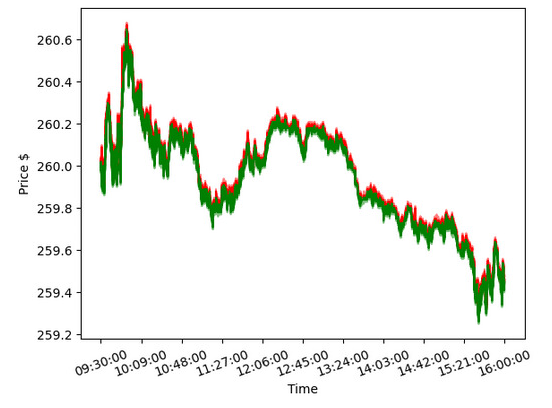}
  \caption{Simulated LOB dynamics}
  \label{fig:sfig1}
\end{subfigure}%
\begin{subfigure}[c]{.32\linewidth}
  \centering
  \includegraphics[width=.9\linewidth]{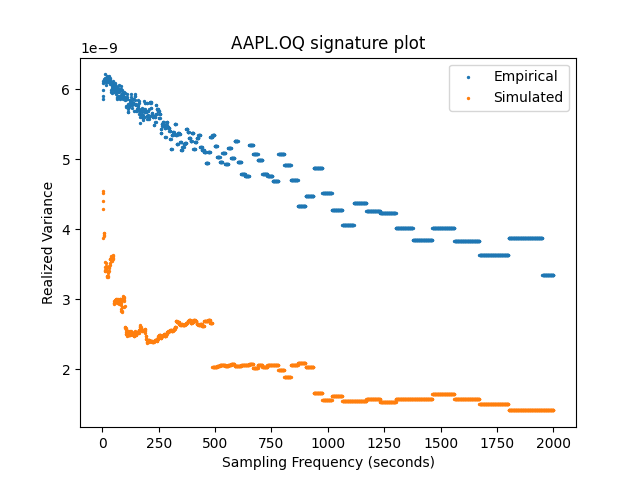}
  \caption{Signature Plot - Simulated vs Empirical}
  \label{fig:sfig2}
\end{subfigure}
\begin{subfigure}[c]{.32\linewidth}
  \centering
  \includegraphics[width=.9\linewidth]{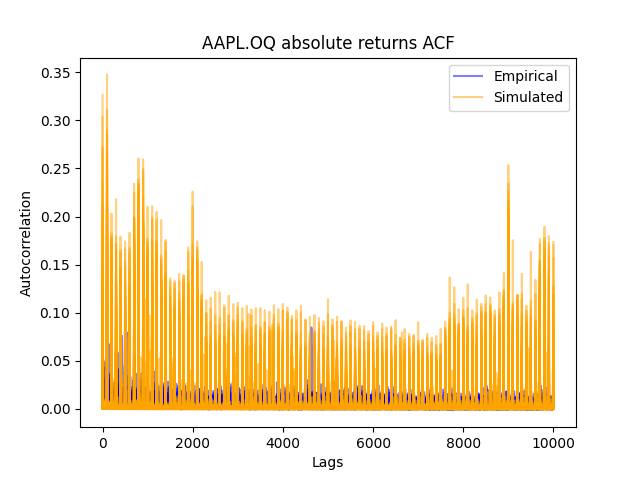}
  \caption{Absolute Returns ACF - Simulated vs Empirical}
  \label{fig:sfig2}
\end{subfigure}
\begin{subfigure}[c]{.32\linewidth}
  \centering
  \includegraphics[width=.9\linewidth]{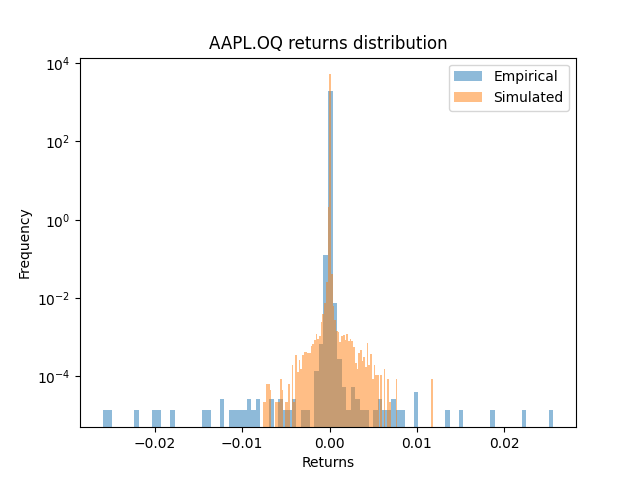}
  \caption{Returns Distribution (log-scale) - Simulated vs Empirical}
  \label{fig:sfig2}
\end{subfigure}
\begin{subfigure}[c]{.32\linewidth}
  \centering
  \includegraphics[width=.9\linewidth]{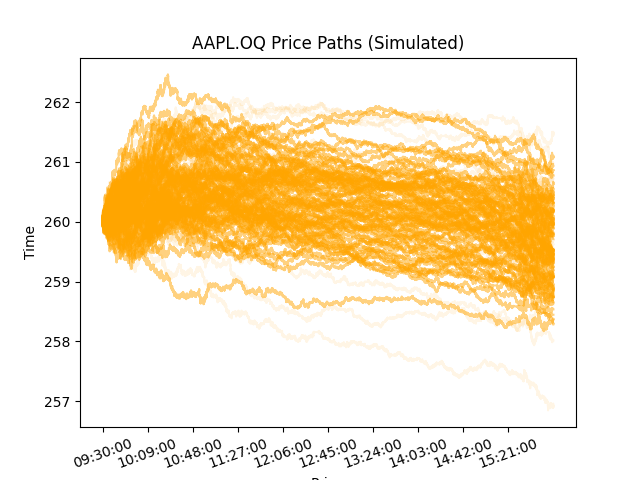}
  \caption{Price Paths - Simulated}
  \label{fig:sfig2}
\end{subfigure}
\begin{subfigure}[c]{.32\linewidth}
  \centering
  \includegraphics[width=.9\linewidth]{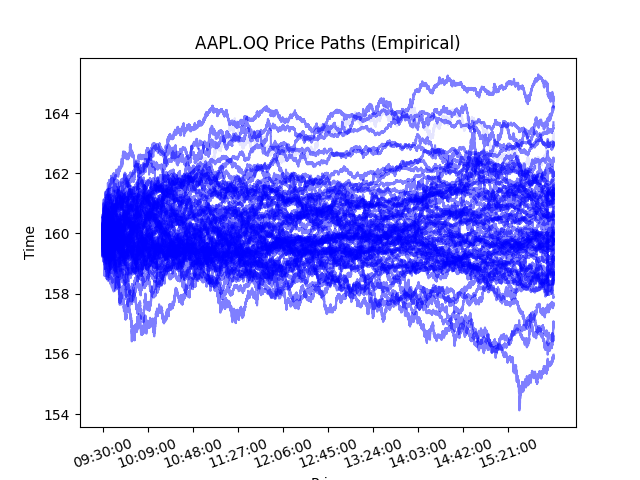}
  \caption{Price Paths - Empirical}
  \label{fig:sfig2}
\end{subfigure}
\begin{subfigure}[c]{.32\linewidth}
  \centering
  \includegraphics[width=.9\linewidth]{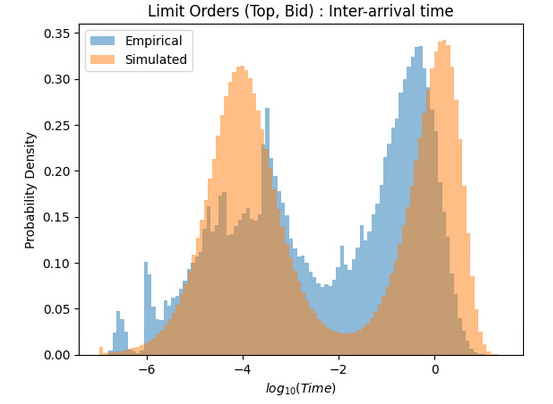}
  \caption{Inter-arrival times - LO (Top, Bid)}
  \label{fig:sfig2}
\end{subfigure}
\begin{subfigure}[c]{.32\linewidth}
  \centering
  \includegraphics[width=.9\linewidth]{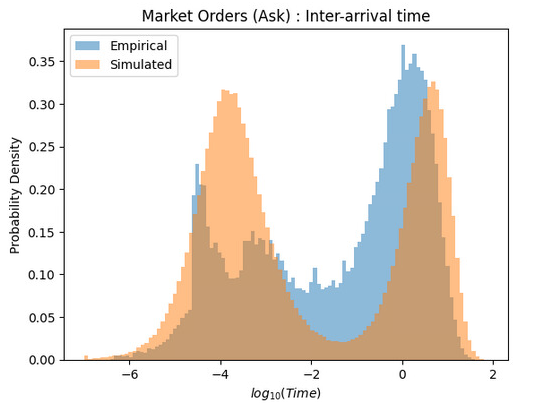}
  \caption{Inter-arrival times - MO (Ask)}
  \label{fig:sfig2}
\end{subfigure}
\begin{subfigure}[c]{.32\linewidth}
  \centering
  \includegraphics[width=.9\linewidth]{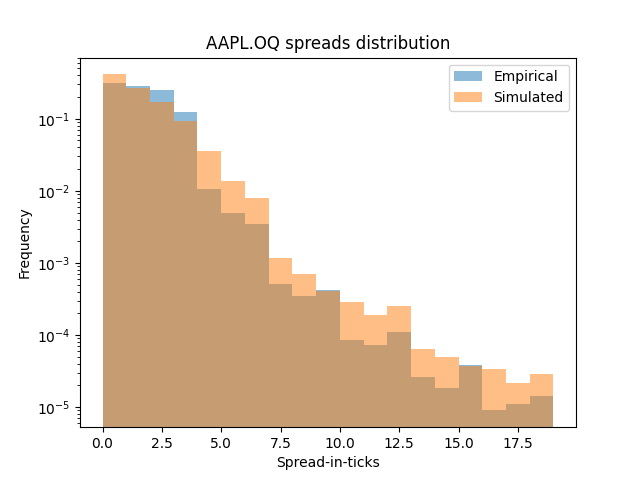}
  \caption{Spread Distribution - Simulated vs Empirical}
  \label{fig:sfig2}
\end{subfigure}
\begin{subfigure}[c]{.32\linewidth}
  \centering
  \includegraphics[width=.9\linewidth]{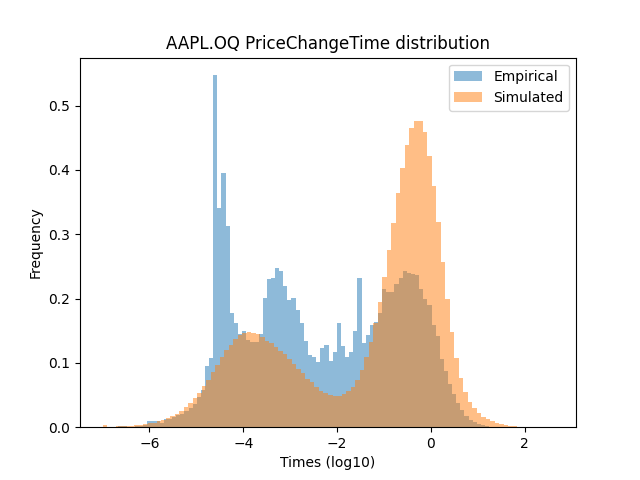}
  \caption{Price Change Time - Simulated vs Empirical}
  \label{fig:sfig2}
\end{subfigure}
\caption{Results (full model)}
\label{fig:qof}
\end{figure}

\begin{figure}
\begin{subfigure}[c]{.32\linewidth}
  \centering
  \includegraphics[width=.95\linewidth]{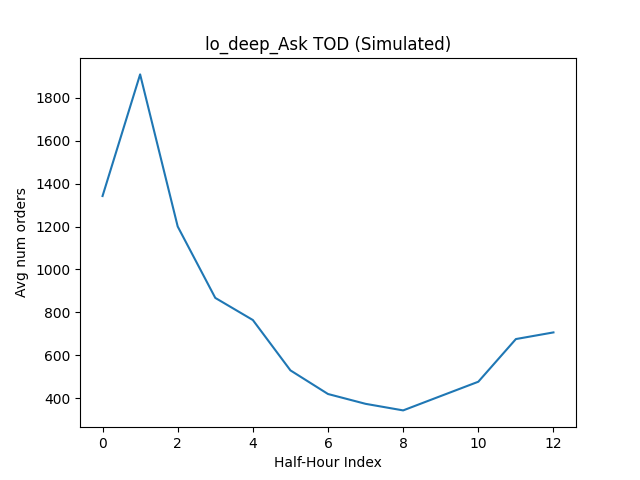}
  \caption{LO(deep, ask) }
  \label{fig:sfig1}
\end{subfigure}%
\begin{subfigure}[c]{.32\linewidth}
  \centering
  \includegraphics[width=.9\linewidth]{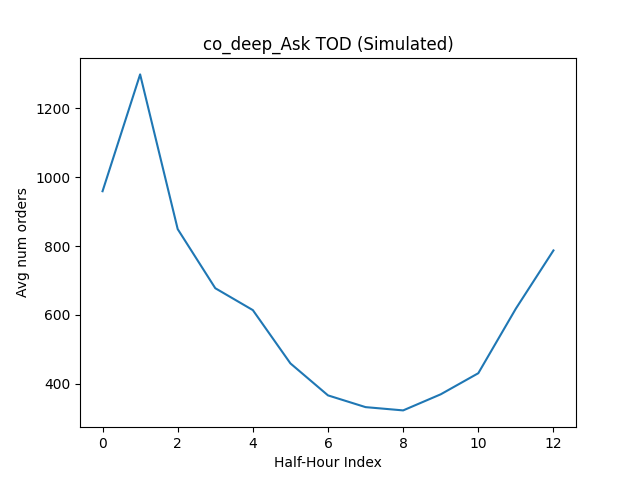}
  \caption{CO(deep, ask) }
  \label{fig:sfig2}
\end{subfigure}
\begin{subfigure}[c]{.32\linewidth}
  \centering
  \includegraphics[width=.9\linewidth]{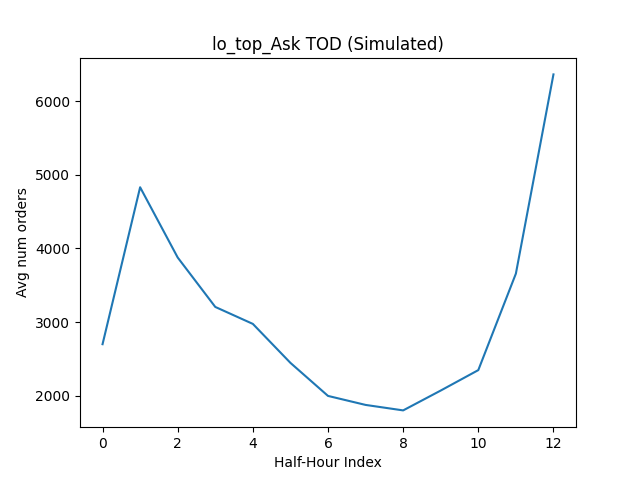}
  \caption{LO(top, ask) }
  \label{fig:sfig2}
\end{subfigure}
\begin{subfigure}[c]{.32\linewidth}
  \centering
  \includegraphics[width=.9\linewidth]{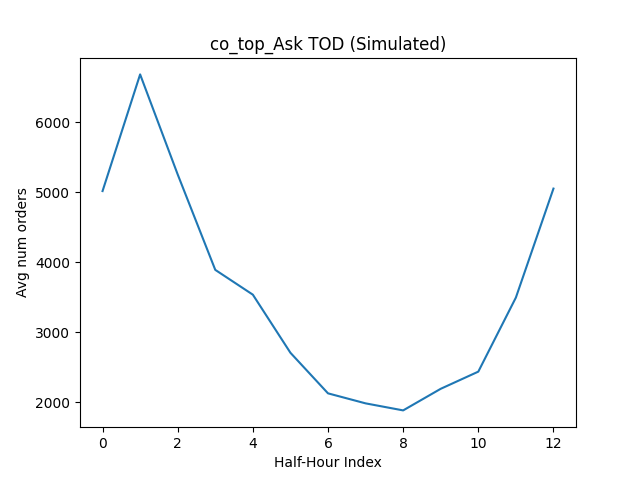}
  \caption{CO(top) }
  \label{fig:sfig2}
\end{subfigure}
\begin{subfigure}[c]{.32\linewidth}
  \centering
  \includegraphics[width=.9\linewidth]{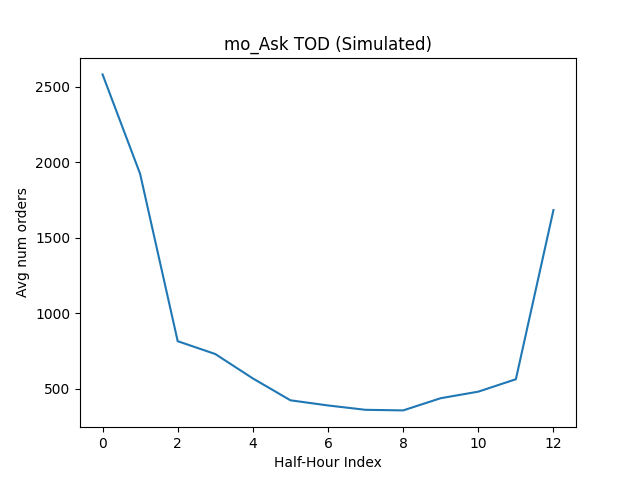}
  \caption{MO(Ask) }
  \label{fig:sfig2}
\end{subfigure}
\begin{subfigure}[c]{.32\linewidth}
  \centering
  \includegraphics[width=.9\linewidth]{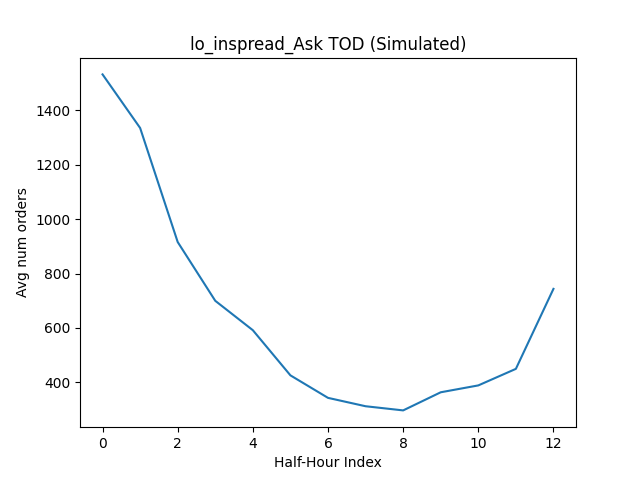}
  \caption{LO(in-spread, Ask) }
  \label{fig:sfig2}
\end{subfigure}
\begin{subfigure}[c]{.32\linewidth}
  \centering
  \includegraphics[width=.9\linewidth]{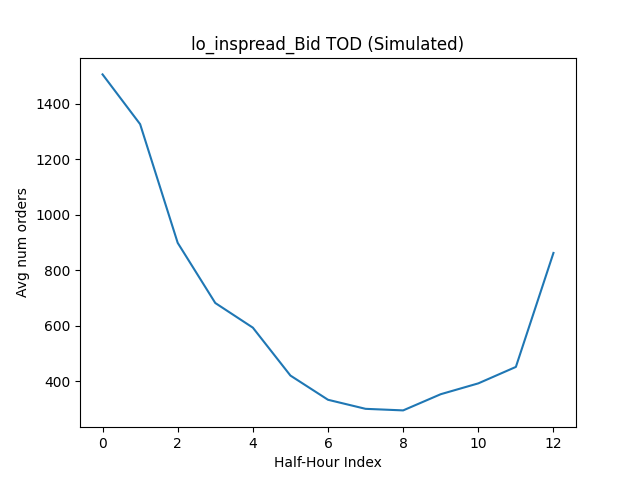}
  \caption{LO(in-spread, Bid) }
  \label{fig:sfig2}
\end{subfigure}
\begin{subfigure}[c]{.32\linewidth}
  \centering
  \includegraphics[width=.9\linewidth]{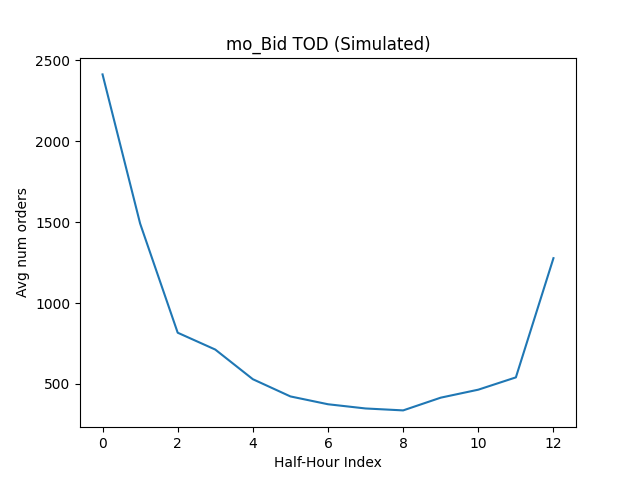}
  \caption{MO(Bid) }
  \label{fig:sfig2}
\end{subfigure}
\begin{subfigure}[c]{.32\linewidth}
  \centering
  \includegraphics[width=.9\linewidth]{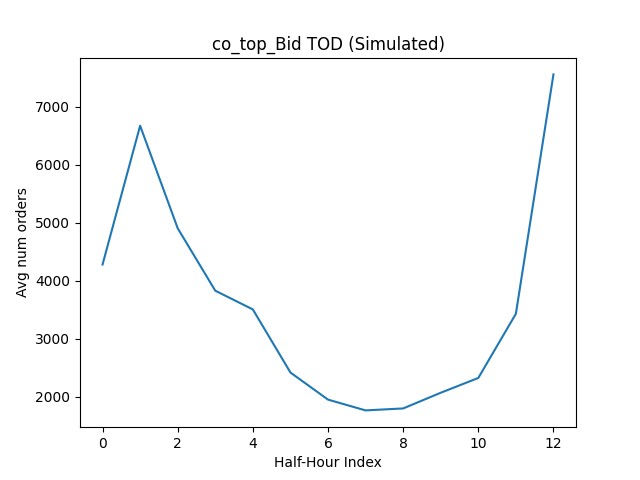}
  \caption{CO(top, Bid) }
  \label{fig:sfig2}
\end{subfigure}
\begin{subfigure}[c]{.32\linewidth}
  \centering
  \includegraphics[width=.9\linewidth]{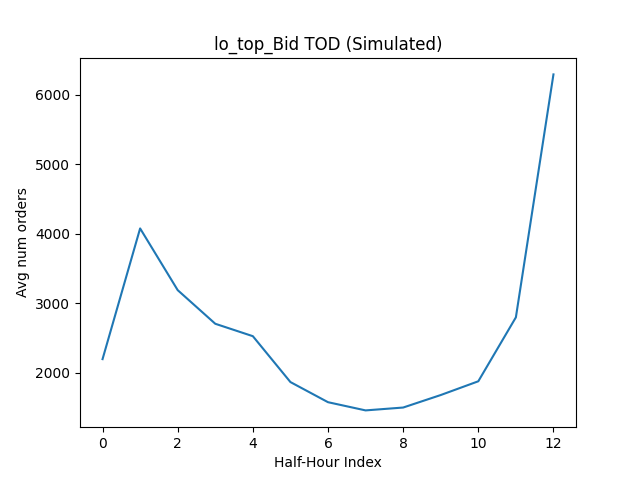}
  \caption{LO(top, Bid) }
  \label{fig:sfig2}
\end{subfigure}
\begin{subfigure}[c]{.32\linewidth}
  \centering
  \includegraphics[width=.9\linewidth]{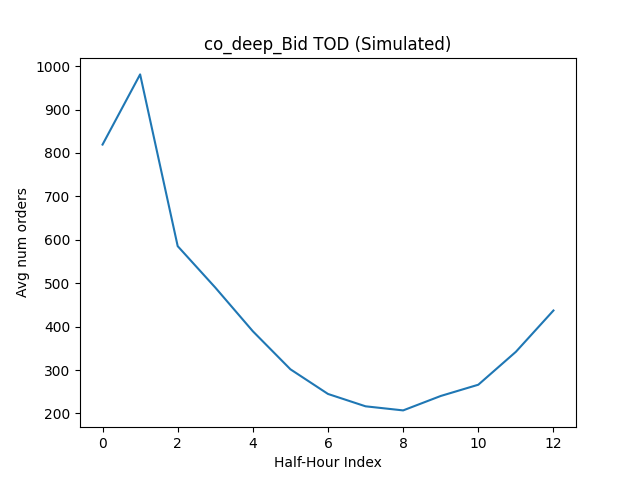}
  \caption{CO(deep, Bid)}
  \label{fig:sfig2}
\end{subfigure}
\begin{subfigure}[c]{.32\linewidth}
  \centering
  \includegraphics[width=.9\linewidth]{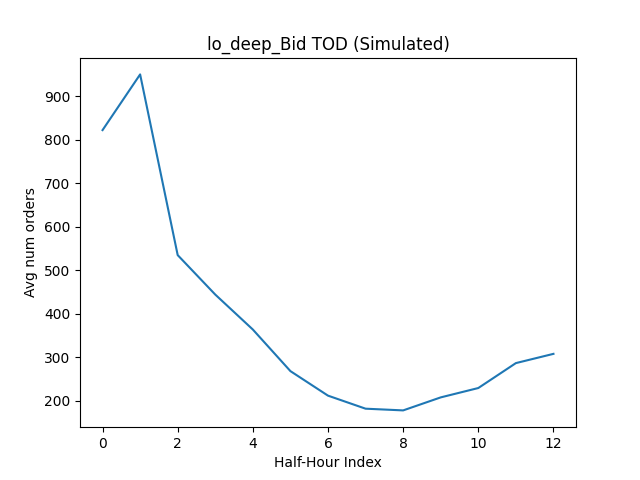}
  \caption{LO(deep, Bid) }
  \label{fig:sfig2}
\end{subfigure}
\caption{Time-of-day shape (simulated)}
\label{fig:todSim}
\end{figure}

We present the quality of fit metrics in Figure \ref{fig:qof}. We perform a realism quality of fit by comparing some stylized facts of the simulated data to the empirical data. We make use of inter-event durations' distributions, price change time distributions, signature plots, distribution of spread and returns, autocorrelation of returns and sample price paths as our set of stylized facts. As we can see from Figure \ref{fig:qof}, the Hawkes Process is able to replicate the shape of the Signature Plot, the presence of long auto-correlations of absolute returns, the long tail of returns' distribution as well as the two peaks in inter-order arrival times quite well. In addition to this, our particular formulation of the Hawkes Process, with its spread control formulation, is able to match the distribution of the empirical spreads quite well.  Figure \ref{fig:todSim} depicts the efficacy of replicating the "U"-shape dynamics of the order book events in our formulation.


\subsection{{Comparison to a similar Poisson Process formulation:}}

{We formulate a 12D Compound Poisson Process with the order intensities of the individual counting processes being Poisson Intensities instead of Hawkes Intensities. The order sizes are sampled from the same distribution as our CHP model. Prop 1 and Prop 2 are enforced by flooring the in-spread order arrival intensity to zero when the spread is 1-tick wide and we multiply the order intensity by the same $f(Q_t)$ time-of-day multiplier. We calibrate the Poisson intensities by the maximum likelihood method and report the quality of fit metrics in Figure \ref{fig:Poisson}. As we can clearly see in the plots, the Poisson model misses the a large mass of the inter-arrival time distribution as well as the price change time distribution. The spread distribution is unrealistic as well and the volatility signature plot is uncharacteristically flat.
\begin{figure}
\begin{subfigure}[c]{.32\linewidth}
  \centering
  \includegraphics[width=.9\linewidth]{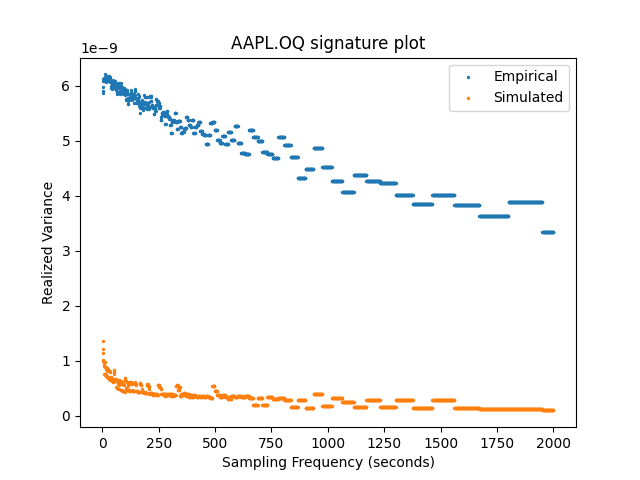}
  \caption{Signature Plot - Simulated vs Empirical}
  \label{fig:sfig2}
\end{subfigure}
\begin{subfigure}[c]{.32\linewidth}
  \centering
  \includegraphics[width=.9\linewidth]{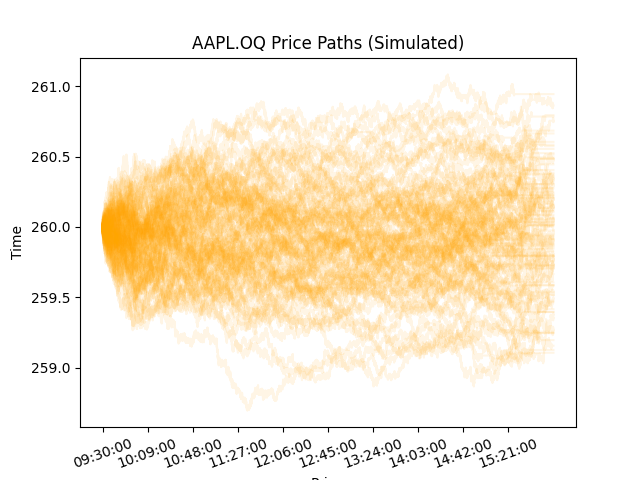}
  \caption{Price Paths - Simulated}
  \label{fig:sfig2}
\end{subfigure}
\begin{subfigure}[c]{.32\linewidth}
  \centering
  \includegraphics[width=.9\linewidth]{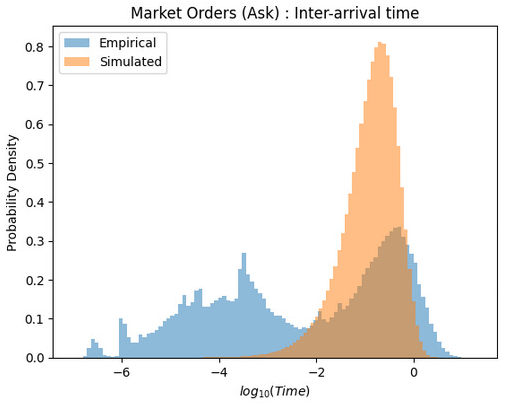}
  \caption{Inter-arrival times - LO (Top, Bid)}
  \label{fig:sfig2}
\end{subfigure}
\begin{subfigure}[c]{.32\linewidth}
  \centering
  \includegraphics[width=.9\linewidth]{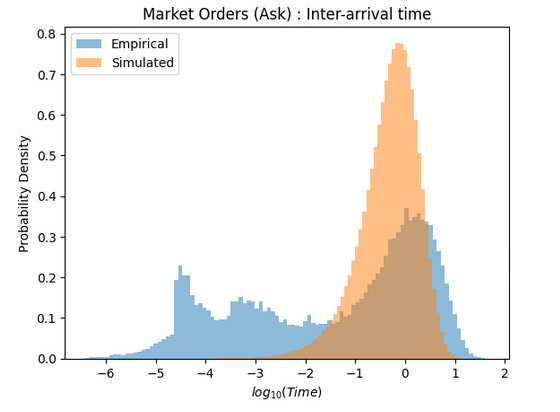}
  \caption{Inter-arrival times - MO (Ask)}
  \label{fig:sfig2}
\end{subfigure}
\begin{subfigure}[c]{.32\linewidth}
  \centering
  \includegraphics[width=.9\linewidth]{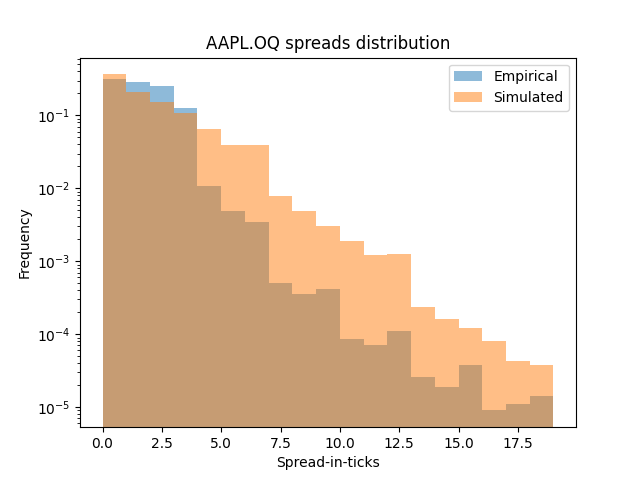}
  \caption{Spread Distribution - Simulated vs Empirical}
  \label{fig:sfig2}
\end{subfigure}
\begin{subfigure}[c]{.32\linewidth}
  \centering
  \includegraphics[width=.9\linewidth]{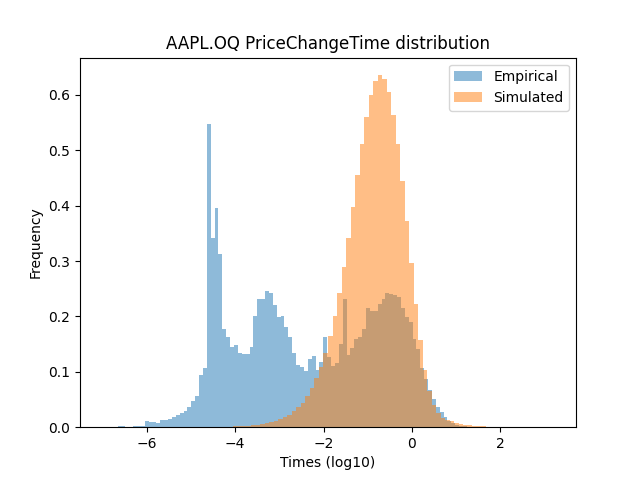}
  \caption{Price Change Time - Simulated vs Empirical}
  \label{fig:sfig2}
\end{subfigure}
\caption{Results (Poisson model)}
\label{fig:Poisson}
\end{figure}
}%
\subsection{{The effect of mis-specifying the order size distribution:}}

{We now show the necessity to model the order sizes' distribution by considering the calibrated kernels with a constant order size (equal to the median order size). We compare the empirical data's price change event times, the volatility signature plot and the spread distribution with the simulated one for the two types of models. As we can clearly see in Figure \ref{fig:qofConsttOrderSize}. As we can see here, our model with the calibrated order size distribution fits much better to the empirical data than the constant order size model - the signature plot and the returns distribution show that the volatility is heavily overestimated (by a magnitude of $\sim100$ as can be seen in the signature plot), the price changes happen much faster as can be seen in the left tail of the price change time plot, and finally the spreads are much wider in general. Our postulate regarding this observation is that price changes depend on not just order arrival times but also on how fast the queue depletes. Therefore, modelling the order sizes will enable for a more realistic price change event times, returns, spreads and volatility in the simulator. 
\begin{figure}
\begin{subfigure}[c]{.32\linewidth}
  \centering
  \includegraphics[width=.9\linewidth]{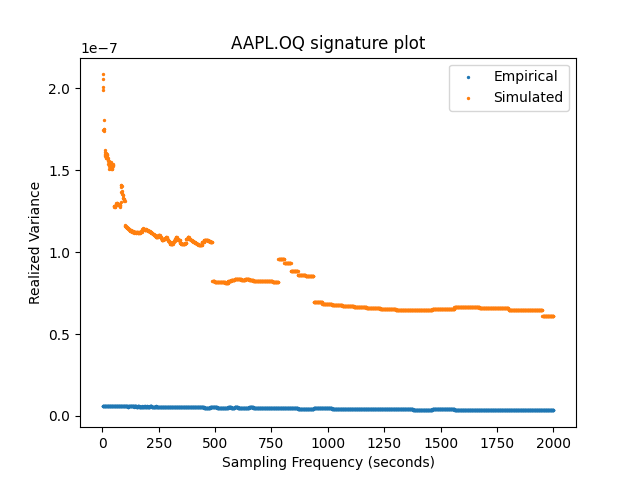}
  \caption{Signature Plot - Simulated vs Empirical}
  \label{fig:sfig2}
\end{subfigure}
\begin{subfigure}[c]{.32\linewidth}
  \centering
  \includegraphics[width=.9\linewidth]{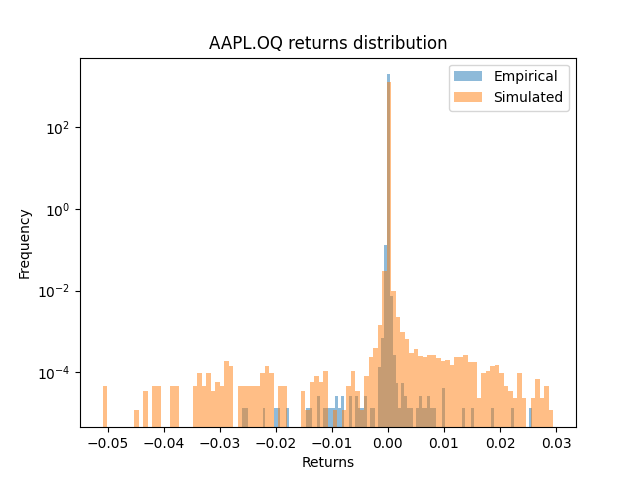}
  \caption{Returns Distribution (log-scale) - Simulated vs Empirical}
  \label{fig:sfig2}
\end{subfigure}
\begin{subfigure}[c]{.32\linewidth}
  \centering
  \includegraphics[width=.9\linewidth]{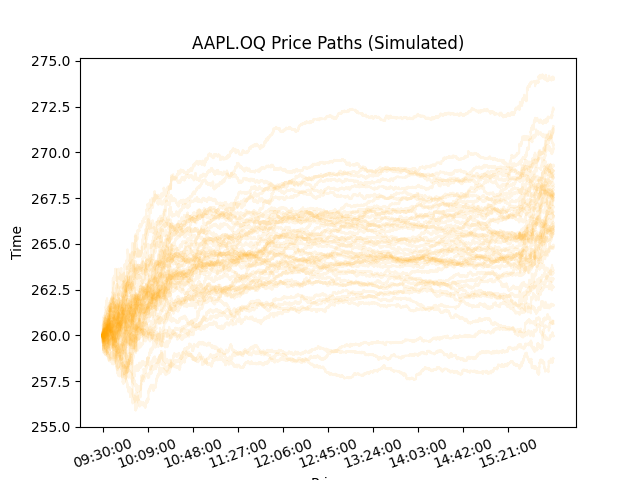}
  \caption{Price Paths - Simulated}
  \label{fig:sfig2}
\end{subfigure}
\begin{subfigure}[c]{.32\linewidth}
  \centering
  \includegraphics[width=.9\linewidth]{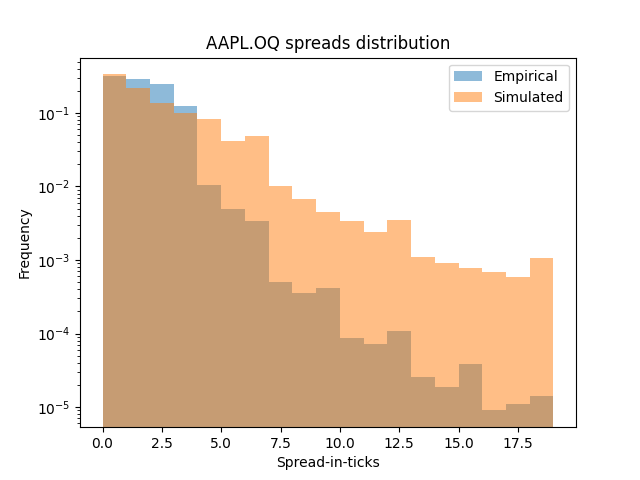}
  \caption{Spread Distribution - Simulated vs Empirical}
  \label{fig:sfig2}
\end{subfigure}
\begin{subfigure}[c]{.32\linewidth}
  \centering
  \includegraphics[width=.9\linewidth]{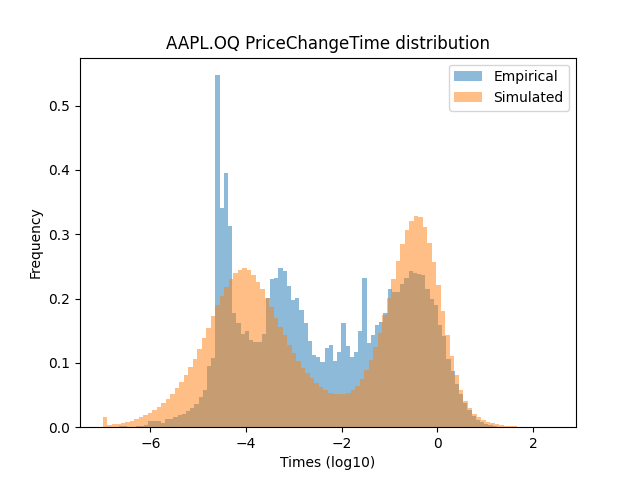}
  \caption{Price Change Time - Simulated vs Empirical}
  \label{fig:sfig2}
\end{subfigure}
\caption{Results (Constant Order Size model)}
\label{fig:qofConsttOrderSize}
\end{figure}
}%
\subsection{{Comparison with Other Hawkes Process Models:}}

{In order to compare our model's performance with the state of the art model, we chose  \citealt{bacry2016estimation}'s 8D Hawkes Process model which has the following types of events: \{Price Changes, Market Orders (no price change), Limit Orders (top, no price change), Cancel Orders (top, no price change\} $\times$ \{ bid, ask \}  . This model takes an alternative approach of modelling the top of the order book than us. We make use of the code made available by their research group\citealt{bacry2017tick} and implement an Expectation-Maximisation fit over the events on the same dataset that we fit our model. We find that the power law kernels again fit the data better than exponential kernels. We infer the power law parameters from the point estimates fitted by the EM Algorithm using the same MLE method we used in our methodology. We then simulate the Hawkes Process using the code provided by the authors in the tick library \citealt{bacry2017tick}. Since the simulator is only able to produce event times and not the queue sizes, spreads or the magnitude of the price change, we only use the inter-event arrival times as the stylized fact for comparison since price-paths, returns and spread distributions are not simulated. We show the results in Figure \ref{fig:bacry}.  
\begin{figure}
\begin{subfigure}[c]{.32\linewidth}
  \centering
  \includegraphics[width=.9\linewidth]{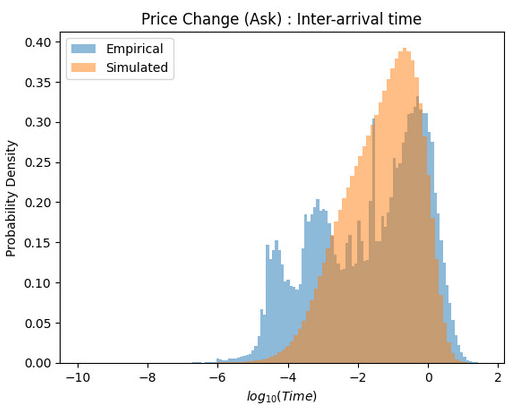}
  \caption{Price Change Time (Ask) - Simulated vs Empirical}
  \label{fig:sfig2}
\end{subfigure}
\begin{subfigure}[c]{.32\linewidth}
  \centering
  \includegraphics[width=.9\linewidth]{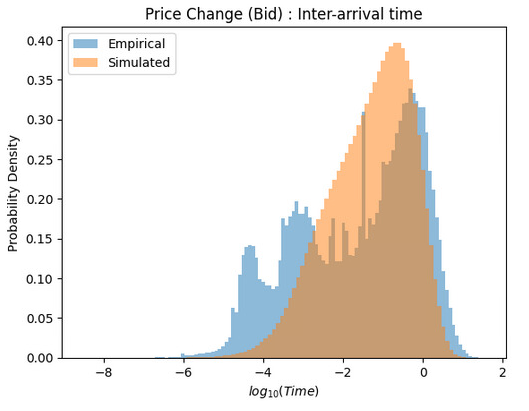}
  \caption{Price Change Time (Bid) - Simulated vs Empirical}
  \label{fig:sfig2}
\end{subfigure}
\begin{subfigure}[c]{.32\linewidth}
  \centering
  \includegraphics[width=.9\linewidth]{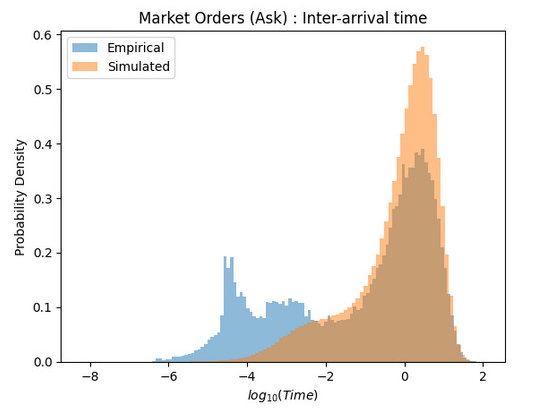}
  \caption{Inter-arrival times - MO (Ask)}
  \label{fig:sfig2}
\end{subfigure}
\begin{subfigure}[c]{.32\linewidth}
  \centering
  \includegraphics[width=.9\linewidth]{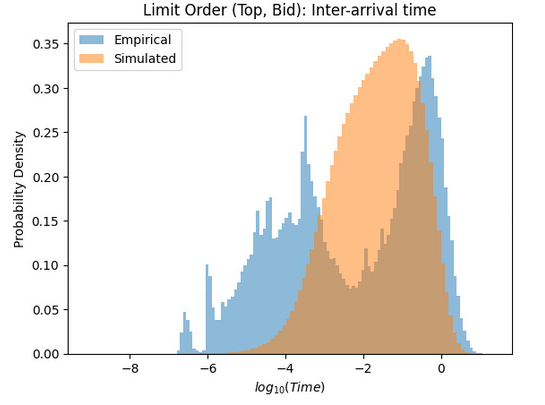}
  \caption{Inter-arrival times - LO (Top, Bid)}
  \label{fig:sfig2}
\end{subfigure}
\caption{Results (Baseline model \citealt{bacry2016estimation})}
\label{fig:bacry}
\end{figure}
As we can see from the simulated distributions, their model is able to capture the right tail of the distribution quite well in all the events we simulate however the model is insufficiently accurate in the high frequency domain (i.e. the left tail).
}%

\subsection{{Market Impact Study:}}

{As noted in \citealt{jain2024limit}, Market Impact (MI) is one of the more crucial characteristics of a realistic order book simulator. We introduce an external agent in the CHP simulator which posts child orders with a quantity $q(Q_T, T, f)$ for a meta order with a quantity $Q_T$, duration of order $T$ and frequency of orders posted $f$. We take inspiration from the setup of \citealt{cont2023limit} in this study and deploy a TWAP strategy (i.e. $q(Q_T, T, f) = \frac{Q_T}{T \times f}$) with varying parameters $Q_T \in \{3000, 6000, 12000, 24000\} \text{ shares}, T \in \{15, 30, 60, 120\} \text{ seconds and } f \in \{1, \frac{1}{3}, \frac{1}{12}, \frac{1}{20}\} \text{ sec}^{-1}$ with order start time fixed. The choice of $Q_T$ is motivated by the fact that average queue size for this stock is found to be around 600 shares, and hence a child order with the most usual configuration in the MI study ($Q_T = 12000, T = 60, f = \frac{1}{3}$) would have a order size equal to the average queue size thus initiating queue depletions. We also conduct a study with varying order start time with other three parameters fixed to $Q_T = 12000, T = 60, f = \frac{1}{3}$. In this way we check for the MI function's sensitivity to each of the 4 parameters we test over. We start the simulation 100 seconds prior to the order start time and observe the mid-price till 120 seconds after the meta order has finished executing. We show simulated mid-price paths in blue, green and yellow respectively for before, during and after the execution of the meta order. We show an average price path in orange. We employ three types of order types for the child orders which are market orders, limit orders (top) and limit orders (deep). Finally, we experiment over both the buy and the sell side of the book. We simulate 100 paths per set of parameters and report some results in Figures \ref{fig:MI1}, \ref{fig:MI2} and \ref{fig:MI3}. 
We now provide an analysis of the observed MI statistics in the simulator. In Figures \ref{fig:MI1} and \ref{fig:MI2}, we focus on MI by Market Orders. We find (Figure \ref{fig:mi_tod}) that trading during the morning has a higher impact than afternoon. This impact again increases near close. This suggests that the time-of-day conditioning of the model parameters propagates to MI as well. Varying the total quantity in the meta-order we find a concave relation between the MI observed and $Q_T$ (Figure \ref{fig:mi_q_fit}). This is inline with the empirically observed concavity in MI with respect to total order quantity. Next testing over the speed of trading, i.e. the percent-of-volume (POV) rate, we vary the time horizon of the meta order $T$ while keeping all other parameters constant. Interestingly we observe a rise in absolute MI $\Delta P$ with increasing $T$ (and decreasing POV rate). We note however a rapid decreasing relation between the rate of change of $\Delta P$ which we denote as $\alpha$ in these experiments. This suggests that the behaviour of $\Delta P$ could be due to the frequency of trading rather than the POV rate. Indeed in Figure \ref{fig:mi_f_fit}, we observe an increase in trading frequency leads to higher $\Delta P$. We also note the price jumps and subsequent relaxation observed in low frequency trading regime (bottom right in Figure \ref{fig:mi_f}). This suggests that our simulator is more sensitive to number of trades than volume of trade for price impact. This is ofcourse due to the assumption we made that order sizes are independent of each other and only the order count excites future orders. This suggests that in this simulator a trading engine might be better off by trading big block trades at a lower frequency instead of using smaller batches which is counter-intuitive to the real world phenomenon of adverse selection. This is a potential missing piece in our simulator. However we again note the concavity of the MI function with $\frac{1}{T}, f$ both. Which implies a concavity of the MI function with respect to POV rate and number of order slices. This is again inline with empirically observed facts. 
In Figure \ref{fig:MI3}, we showcase the MI of Limit Orders at various depths in the order book. We note the concavity seen in MI for limit orders at the top of the book in Figure \ref{fig:mi_lo_top_fit}. We also observe that limit orders have a much weaker and noisier (but non-zero nonetheless) MI on the LOB than the MOs. The MI weakens to being indistinguishable from noise as we go deeper in the book but at the cost of fill probability. This observation also is inline with observations from empirical data. 
\begin{figure}
\begin{subfigure}[c]{.95\linewidth}
  \centering
  \includegraphics[width=.9\linewidth]{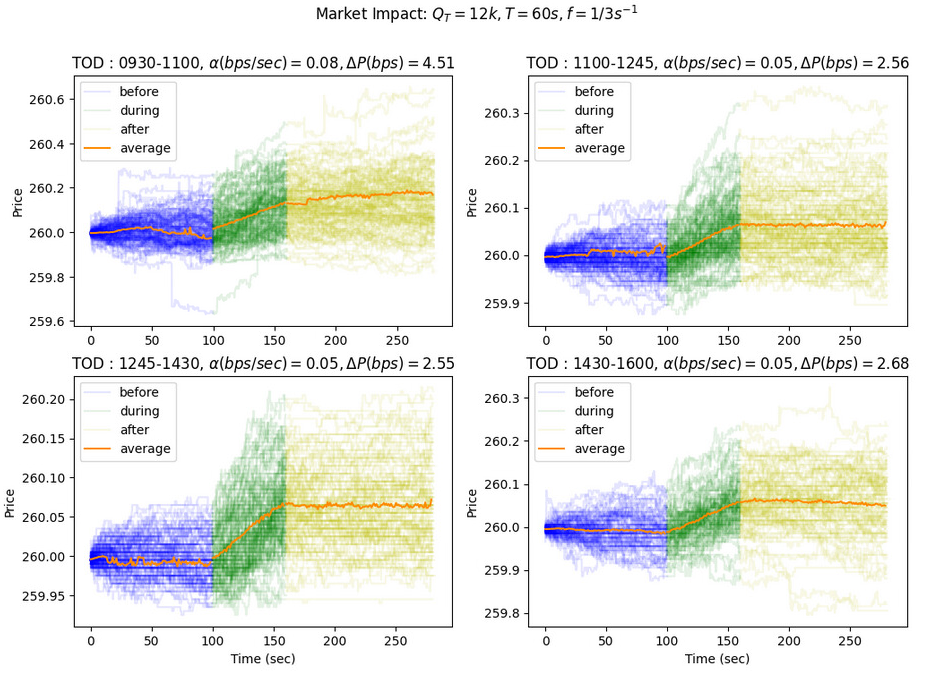}
  \caption{By TOD (MO, Buy)}
  \label{fig:mi_tod}
\end{subfigure}
\begin{subfigure}[c]{.45\linewidth}
  \centering
  \includegraphics[width=.9\linewidth]{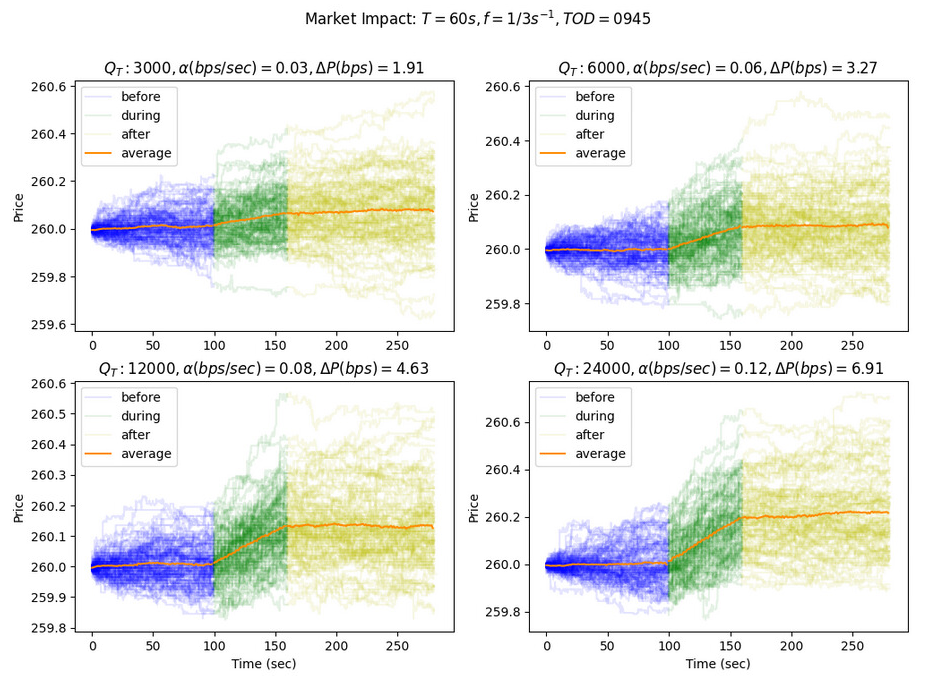}
  \caption{By $Q_T$ (MO, Buy)}
  \label{fig:mi_q}
\end{subfigure}
\begin{subfigure}[c]{.45\linewidth}
  \centering
  \includegraphics[width=.9\linewidth]{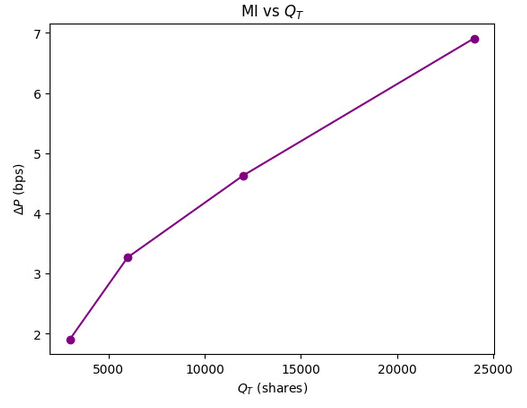}
  \caption{MI w.r.t. $Q_T$ (MO, Buy)}
  \label{fig:mi_q_fit}
\end{subfigure}
\caption{Market Impact Study (Market Orders)}
\label{fig:MI1}
\end{figure}
\begin{figure}
\begin{subfigure}[c]{.45\linewidth}
  \centering
  \includegraphics[width=.9\linewidth]{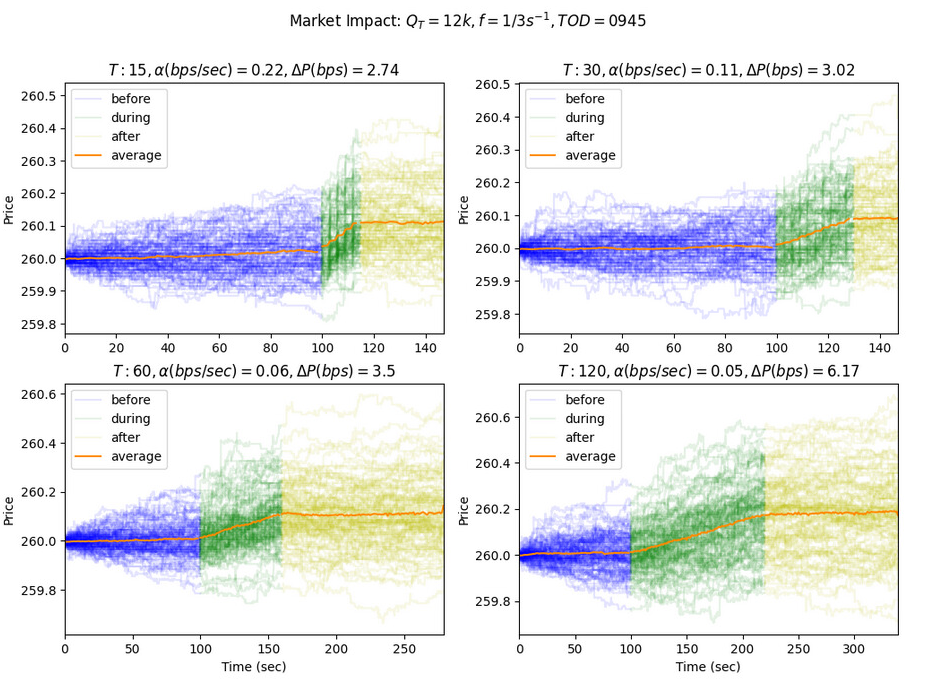}
  \caption{By $T$ (MO, Buy)}
  \label{fig:mi_t}
\end{subfigure}
\begin{subfigure}[c]{.45\linewidth}
  \centering
  \includegraphics[width=.9\linewidth]{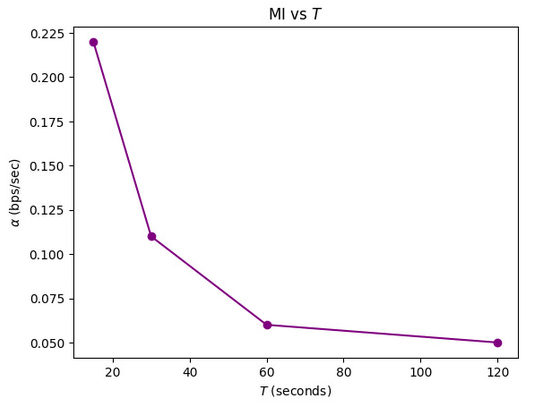}
  \caption{MI w.r.t. $T$ (MO, Buy)}
  \label{fig:mi_t_fit}
\end{subfigure}
\begin{subfigure}[c]{.45\linewidth}
  \centering
  \includegraphics[width=.9\linewidth]{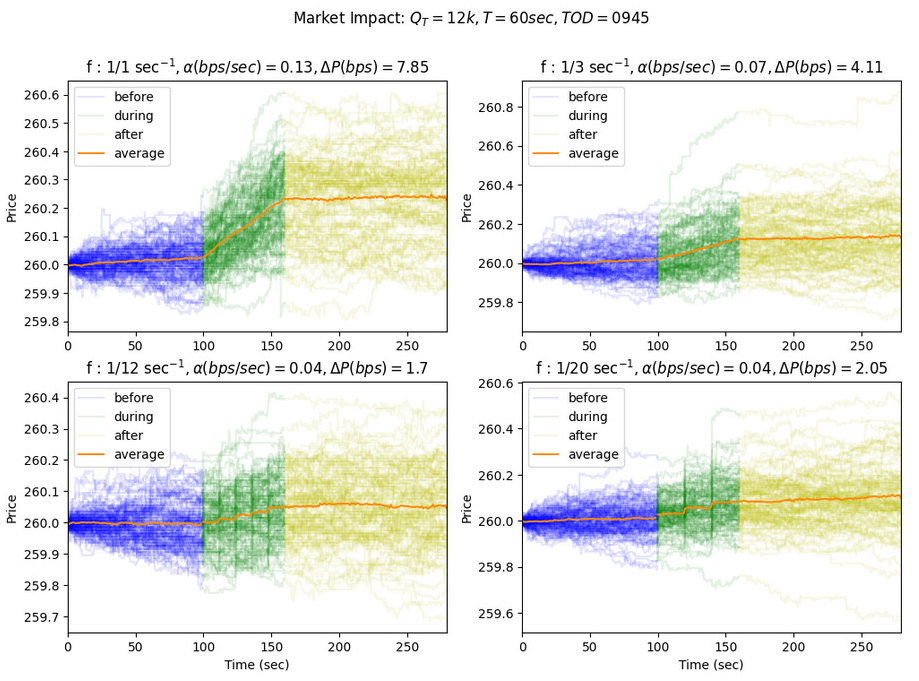}
  \caption{By $f$ (MO, Buy)}
  \label{fig:mi_f}
\end{subfigure}
\begin{subfigure}[c]{.45\linewidth}
  \centering
  \includegraphics[width=.9\linewidth]{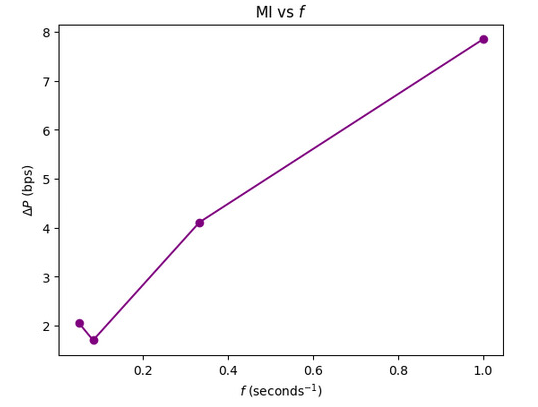}
  \caption{MI w.r.t. $f$ (MO, Buy)}
  \label{fig:mi_f_fit}
\end{subfigure}
\caption{Market Impact Study (Market Orders, continued)}
\label{fig:MI2}
\end{figure}

\begin{figure}
\centering
\begin{subfigure}[c]{.45\linewidth}
  \centering
  \includegraphics[width=0.9\linewidth]{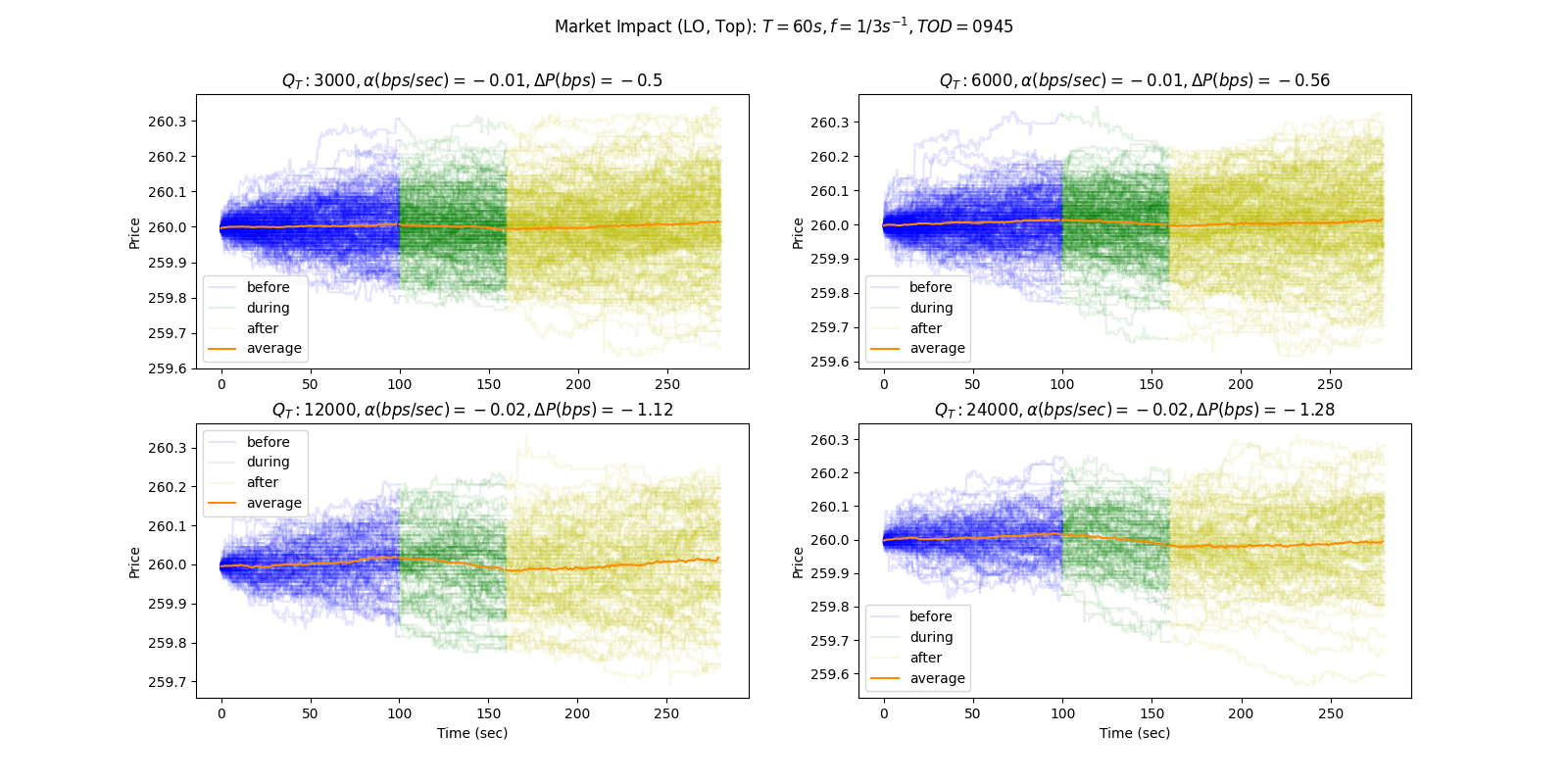}
  \caption{By $Q_T$ (LO, Top, Sell)}
  \label{fig:mi_lo_top}
\end{subfigure}
\begin{subfigure}[c]{.45\linewidth}
  \centering
  \includegraphics[width=.9\linewidth]{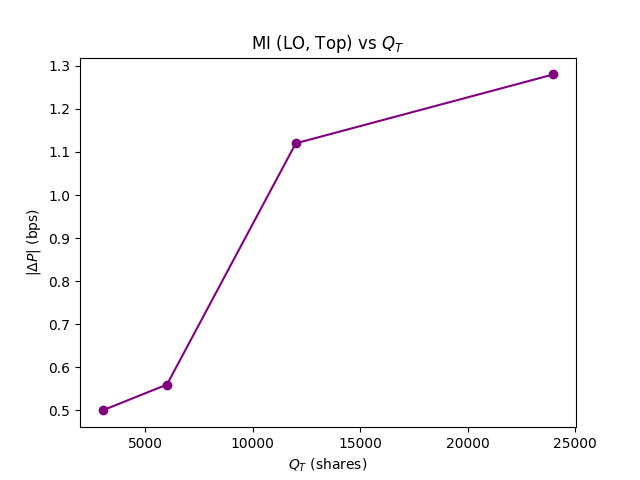}
  \caption{MI w.r.t. $Q_T$ (LO, Top, Sell)}
  \label{fig:mi_lo_top_fit}
\end{subfigure}
\begin{subfigure}[c]{.45\linewidth}
  \centering
  \includegraphics[width=.9\linewidth]{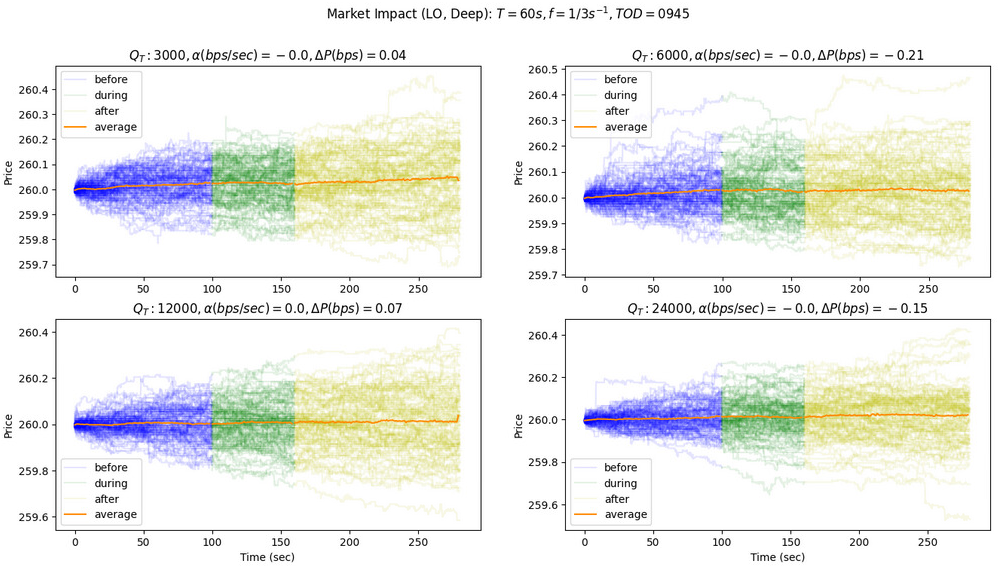}
  \caption{By $Q_T$ (LO, Deep, Buy)}
  \label{fig:mi_lo_deep}
\end{subfigure}
\begin{subfigure}[c]{.45\linewidth}
  \centering
  \includegraphics[width=.9\linewidth]{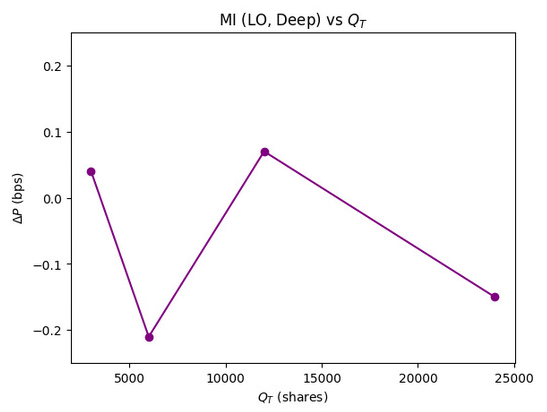}
  \caption{MI w.r.t. $Q_T$ (LO, Deep, Buy)}
  \label{fig:mi_lo_deep_fit}
\end{subfigure}
\caption{Market Impact Study (Limit Orders)}
\label{fig:MI3}
\end{figure}
}%
\section{Conclusion \& Future Work}

We tackle the problem of simulating a realistic order book by using the Compound Hawkes Process. We particularly focus on building a simulator which is realistic in its order sizes and exhibits the mechanical constraints that a real order book has such as non-negative spreads and positive intensities. We further condition the intensities to be dependent on time-of-day in order for the simulator to not be ignorant of the increase in trading intensities around open and close auctions.We present a number of stylized facts to test the model's efficacy against empirical observations and we compare it our model's empirical facts to several baselines.

In this formulation of the Hawkes Process, we assume nothing about the Price Process of the equity and rather let it rather emerge as a jump process by modest assumptions about the order flow itself. Our contributions can be enlisted as follows:

\begin{enumerate}
    \item We propose a novel formulation of the Hawkes Process to maintain positive spread.
    \item We augment the non-parameteric methods proposed in \citealt{kirchner2017estimation} to work with slow decaying kernels and to be more stable.
    \item We make use of calibrated distributions for sampling the order sizes of the events rather than assuming unit order-size.
    \item We formulate the in-spread order arrivals as a function of the current spread - utilizing the well known fact that spreads are mean-reverting.
    \item We perform a MI study on this simulator along with testing against several baselines and empirical data by making use of a number of popular stylized facts.
\end{enumerate}

Future directions of research includes testing this model on more equities, particularly tackle the problem of small-tick stocks which have a lower amount of information at the top two levels of the LOB which our model assumes. Though our formulation theoretically can work with small-tick stocks as well but since it limits its modelling to the top of the book, it will be unsuitable for small-tick stocks. \textcolor{black}{By large, small-tick stocks violate a number of assumptions that we make in this model of the LOB - the LOB is not dense in small-tick stocks, neither is the price improvement done uniformly at 1-tick from the best, and finally there exists quite a few deeper levels than the top two which affect the LOB dynamics. To generalize this model for such stocks, future research should be focused on relaxing these assumptions, one possible avenue is modelling more levels of the LOB in the Hawkes process albeit at the cost of computational time and accuracy.} Further, the number of dimensions of this Hawkes Process is quite high, future research could be focused on simulating the order book with lesser number of dimensions. 

\appendix

\section{Previous Order Sizes do not impact Future Order Arrival Rates}

\begin{figure*}
\begin{subfigure}[c]{.49\linewidth}
  \centering
  \includegraphics[width=.7\linewidth]{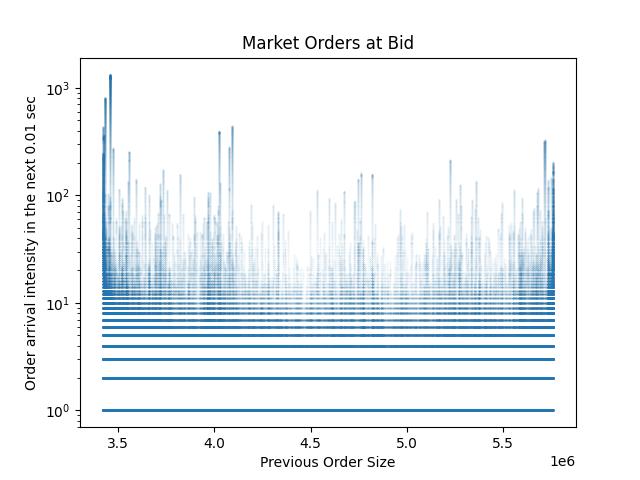}
  \caption{Previous Order Size vs Future Arrival Rates }
  \label{fig:sizeVsArrival}
\end{subfigure}%
\begin{subfigure}[c]{.49\linewidth}
  \centering
  \includegraphics[width=.7\linewidth]{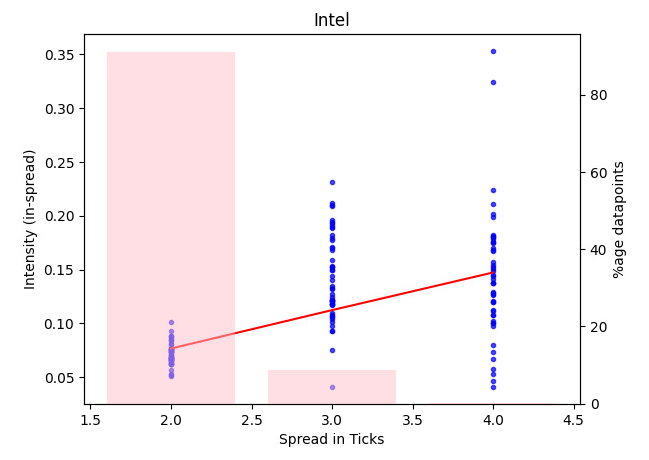}
  \caption{In-Spread Arrival Rates vs Current Spread : INTC.OQ}
  \label{fig:spread3}
\end{subfigure}
\begin{subfigure}[c]{.49\linewidth}
  \centering
  \includegraphics[width=.7\linewidth]{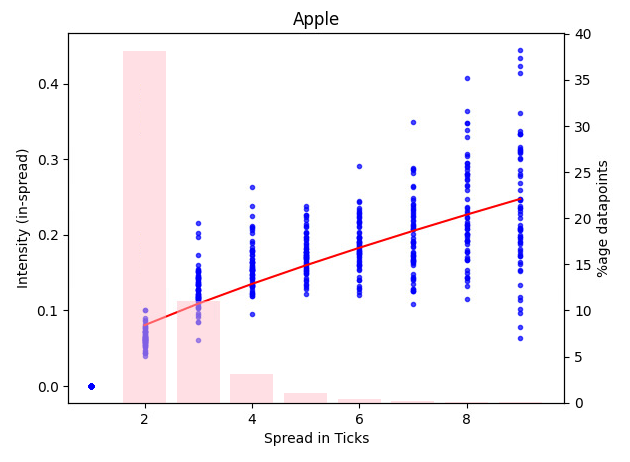}
  \caption{In-Spread Arrival Rates vs Current Spread : AAPL.OQ}
  \label{fig:spread}
\end{subfigure}
\begin{subfigure}[c]{.49\linewidth}
  \centering
  \includegraphics[width=.7\linewidth]{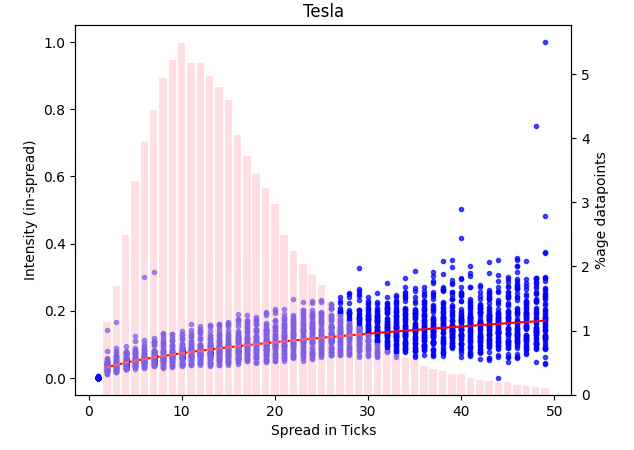}
  \caption{In-Spread Arrival Rates vs Current Spread : TSLA.OQ}
  \label{fig:spread4}
\end{subfigure}
\begin{subfigure}[c]{.49\linewidth}
  \centering
  \includegraphics[width=.7\linewidth]{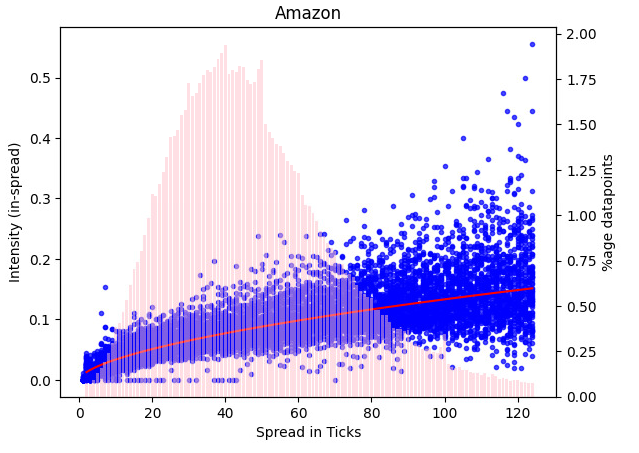}
  \caption{In-Spread Arrival Rates vs Current Spread : AMZN.OQ}
  \label{fig:spread2}
\end{subfigure}
\caption{Appendix}
\label{fig:app}
\end{figure*}

We calculate the next order intensity in our dataset for a past order size by counting the number of future events in a window of $t = 0.01 $sec. The joint scatter plot for Market Orders at Bid is shown in Figure \ref{fig:sizeVsArrival}. Qualitatively the distribution looks to be quite uniform. We make use of the Hoeffding Independence Test to calculate the distance between the observed joint distribution of these two random variables and the distribution if they were independent. The test statistic is observed to be $0.00068$ which is sufficiently low for us to conclude, albeit with weak evidence, that the two variables are independent. We observe similar scatter plots and Hoeffding statistics for all other events. 

\section{In-Spread order intensity depends on current spread}

We plot the order intensity against the current spread-in-ticks in empirical data (3 months data) in Figure \ref{fig:spread}. Here we show a scatter plot since the intensities are approximated by the number of in-spread orders in the next 0.01 seconds and therefore are random. We show the distribution of data points in the translucent red bars. We fit a linear regressor on log-log transformation of these data points (excluding 0 and 1 spread) and find the best exponent to be $0.7479$. The red line in the plot shows the fitted line. We observe an $R^2$ of $0.85$ for this regression. Similar figures for other stocks are also shown in Figures \ref{fig:spread2}, \ref{fig:spread3} , \ref{fig:spread4}.

\section{Results for some more stocks:}

\begin{figure*}
\centering
\begin{subfigure}[c]{.48\linewidth}
\includegraphics[width=.95\linewidth]{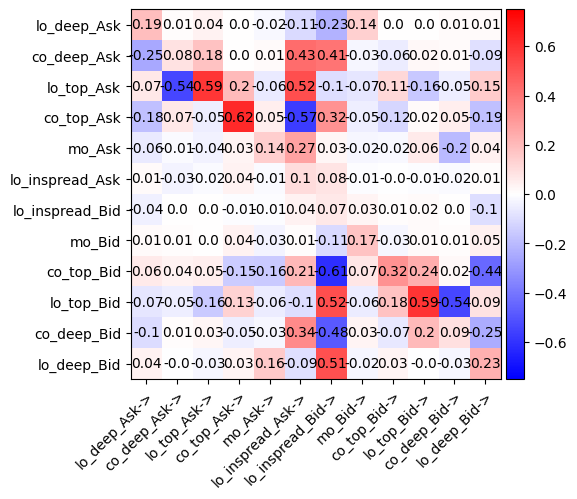}
\hfill
\caption{Norms of Kernel: INTC.OQ}
\label{fig:intcNorms}
\end{subfigure}
\begin{subfigure}[c]{0.48\textwidth}
\includegraphics[width=0.95\linewidth]{{images/AAPL.OQ_kernelNormMatrix}.png}
\hfill
\caption{Norms of Kernel: AAPL.OQ}
\label{fig:kernelNorms}
\end{subfigure}
\begin{subfigure}[c]{.48\linewidth}
\includegraphics[width=0.95\linewidth]{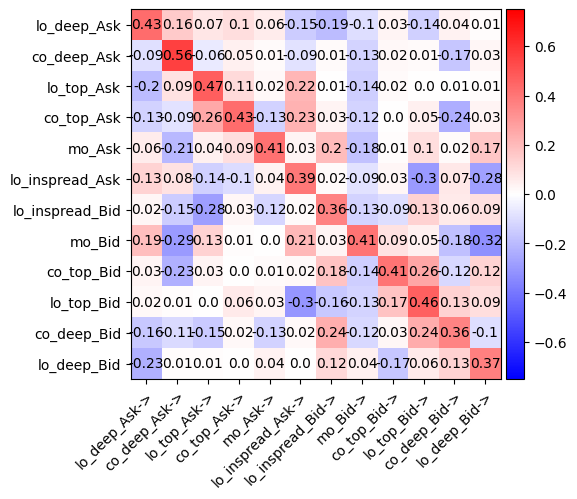}

\caption{Norms of Kernel: TSLA.OQ}
\label{fig:tslaNorms}
\end{subfigure}
\begin{subfigure}[c]{.48\linewidth}
\includegraphics[width=0.95\linewidth]{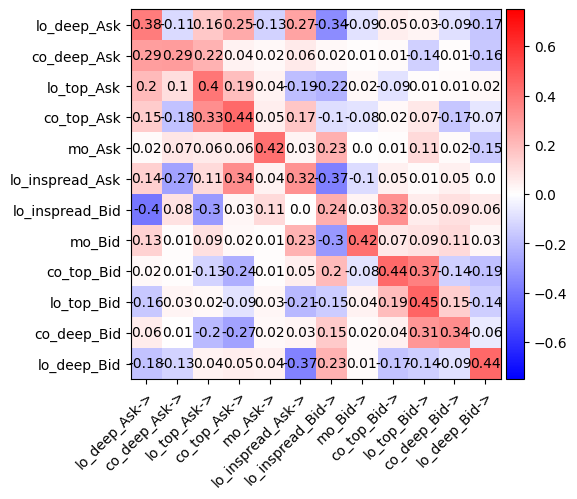}
\hfill
\caption{Norms of Kernel: AMZN.OQ}
\label{fig:tslaNorms}
\end{subfigure}
\caption{Norms of Kernels}
\label{fig:norms}
\end{figure*}

\begin{figure*}
\begin{subfigure}[c]{.32\linewidth}
  \centering
  \includegraphics[width=.95\linewidth]{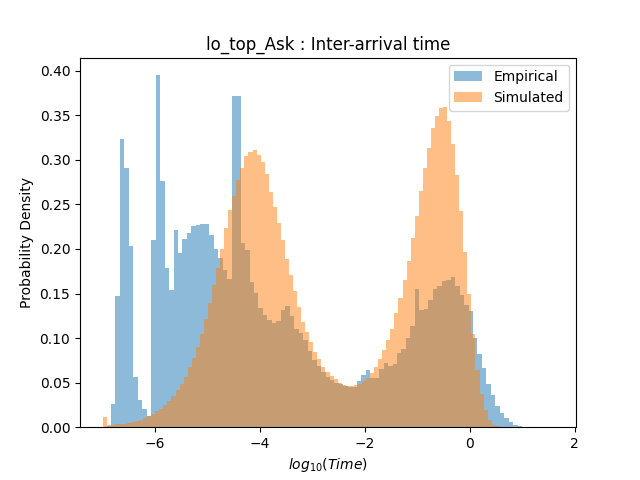}
  \caption{INTC.OQ - LO Top (Ask)}
  \label{fig:sfig1}
\end{subfigure}%
\begin{subfigure}[c]{.32\linewidth}
  \centering
  \includegraphics[width=.9\linewidth]{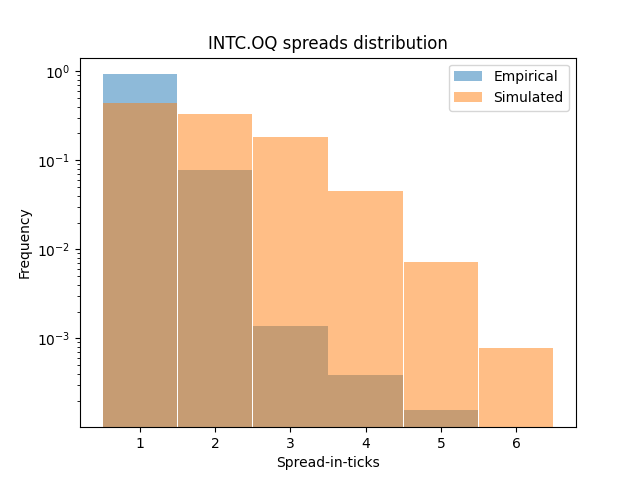}
  \caption{INTC.OQ - Spread Distribution}
  \label{fig:sfig2}
\end{subfigure}
\begin{subfigure}[c]{.32\linewidth}
  \centering
  \includegraphics[width=.9\linewidth]{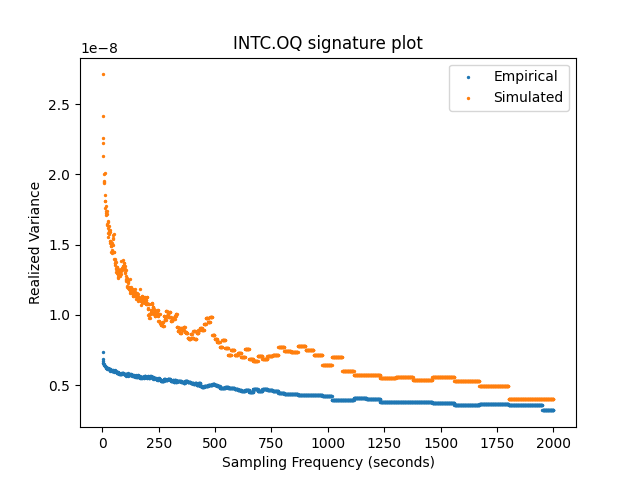}
  \caption{INTC.OQ - Signature Plot}
  \label{fig:sfig2}
\end{subfigure}
\begin{subfigure}[c]{.32\linewidth}
  \centering
  \includegraphics[width=.95\linewidth]{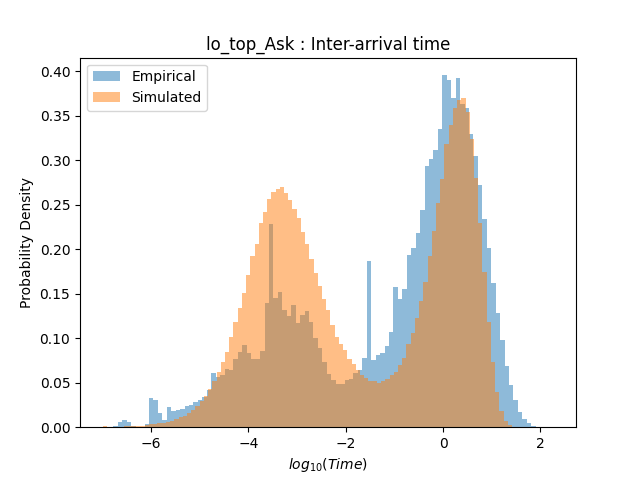}
  \caption{TSLA.OQ - LO Top (Ask)}
  \label{fig:sfig1}
\end{subfigure}%
\begin{subfigure}[c]{.32\linewidth}
  \centering
  \includegraphics[width=.9\linewidth]{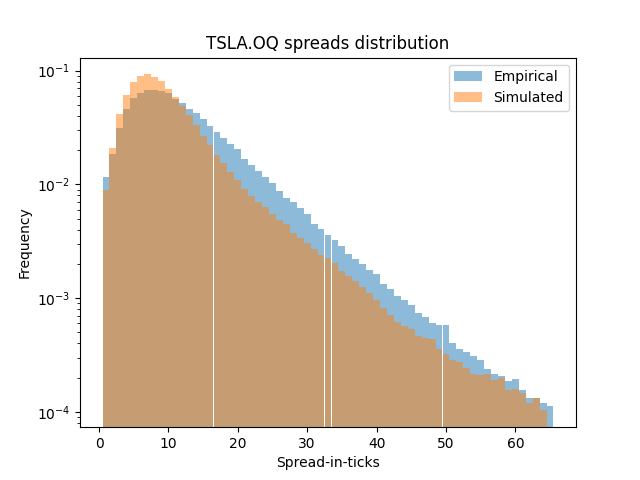}
  \caption{TSLA.OQ - Spread Distribution}
  \label{fig:sfig2}
\end{subfigure}
\begin{subfigure}[c]{.32\linewidth}
  \centering
  \includegraphics[width=.9\linewidth]{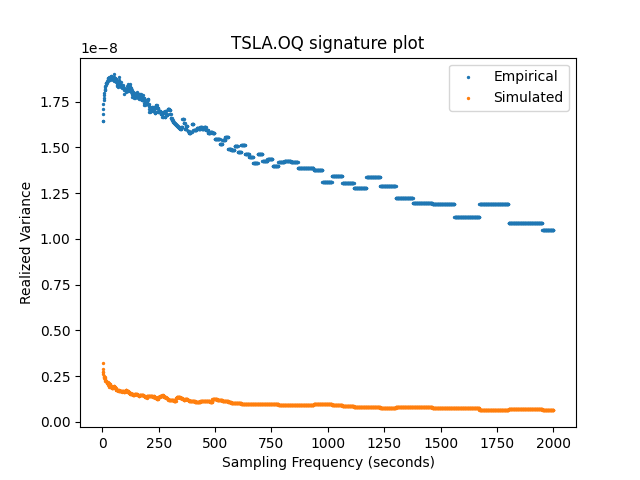}
  \caption{TSLA.OQ - Signature Plot}
  \label{fig:sfig2}
\end{subfigure}
\begin{subfigure}[c]{.32\linewidth}
  \centering
  \includegraphics[width=.95\linewidth]{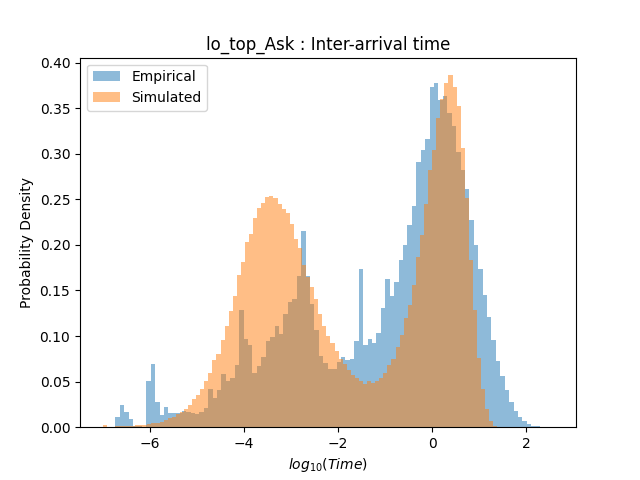}
  \caption{AMZN.OQ - LO Top (Ask)}
  \label{fig:sfig1}
\end{subfigure}%
\begin{subfigure}[c]{.32\linewidth}
  \centering
  \includegraphics[width=.9\linewidth]{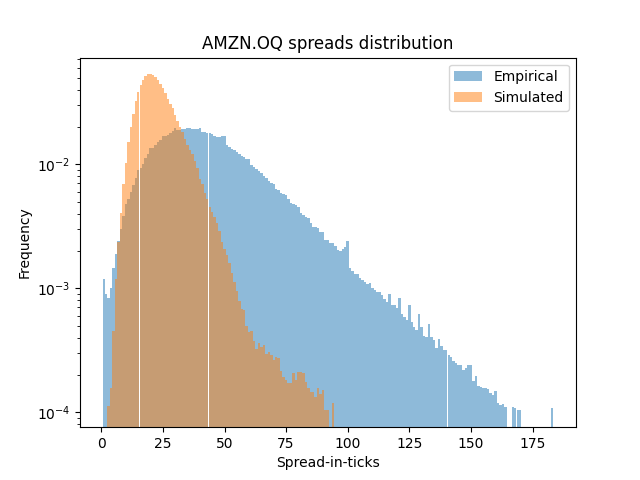}
  \caption{AMZN.OQ - Spread Distribution}
  \label{fig:sfig2}
\end{subfigure}
\begin{subfigure}[c]{.32\linewidth}
  \centering
  \includegraphics[width=.9\linewidth]{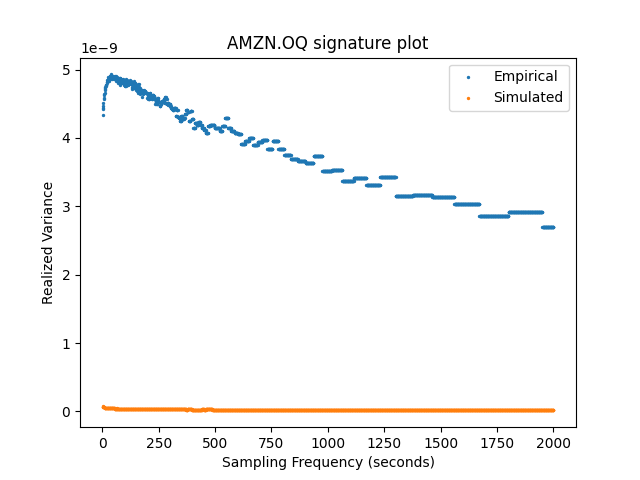}
  \caption{AMZN.OQ - Signature Plot}
  \label{fig:sfig2}
\end{subfigure}
\begin{subfigure}[c]{.32\linewidth}
  \centering
  \includegraphics[width=.9\linewidth]{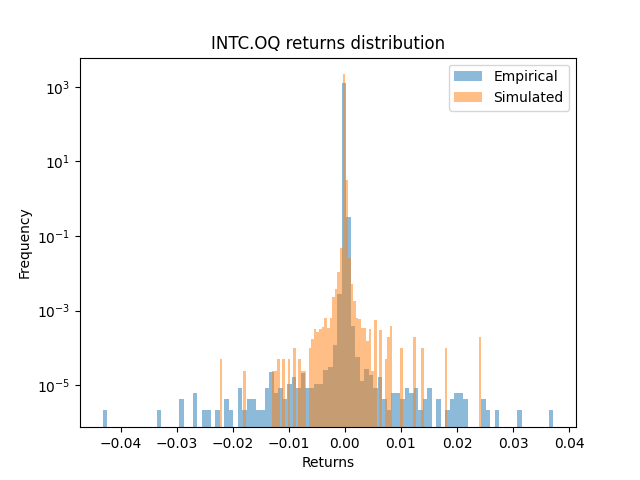}
  \caption{INTC.OQ - Returns Distribution}
  \label{fig:sfig2}
\end{subfigure}
\begin{subfigure}[c]{.32\linewidth}
  \centering
  \includegraphics[width=.9\linewidth]{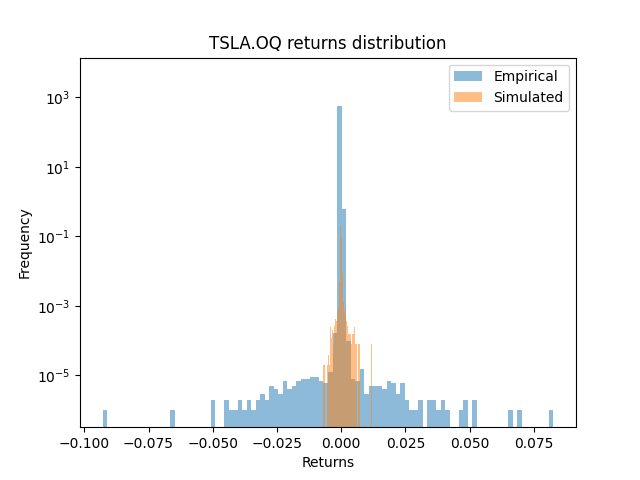}
  \caption{TSLA.OQ - Returns Distribution}
  \label{fig:sfig2}
\end{subfigure}
\begin{subfigure}[c]{.32\linewidth}
  \centering
  \includegraphics[width=.95\linewidth]{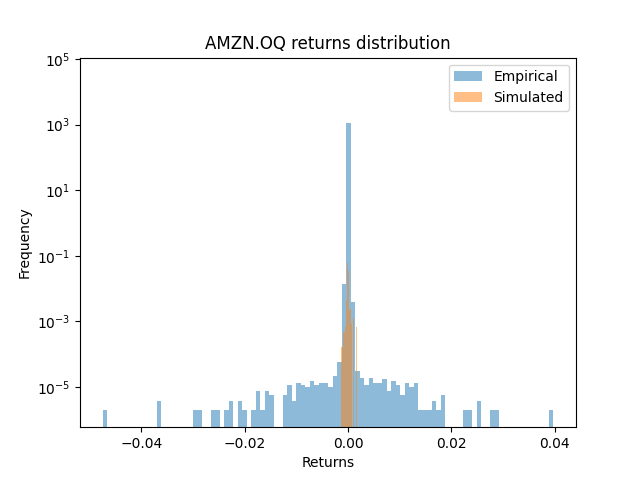}
  \caption{AMZN.OQ - Returns Distribution}
  \label{fig:sfig1}
\end{subfigure}%
\caption{Results (full model - other stocks)}
\label{fig:qof2}
\end{figure*}

In addition to Apple Inc. (Medium-Tick), we also calibrate and simulate the 12D Hawkes Process for Intel (Large-Tick), Tesla (Small-Tick) and Amazon (Very Small-Tick) stocks. In Figure \ref{fig:norms} we show the calibration results for the three stocks. Finally we show the stylized facts comparison between the model and empirical data in Figure \ref{fig:qof2}. For Intel, we see that the Hawkes Process model is not able to capture the high frequency domain of the order book events (Figure \ref{fig:qof2}a.) well and the spread is higher than the observed (Figure \ref{fig:qof2}b.). Tesla and Amazon have relatively better fit in order book events inter-arrival times (Figure \ref{fig:qof2} d. and g. respectively) however the spread distribution is underestimated in the Tesla model and severly underestimated in the Amazon model. Finally the signature plots (Figures \ref{fig:qof2} c., f. and i.) show that while the Intel volatility is slightly overestimated however the scale is similar (like Apple in the Figure \ref{fig:qof}), Tesla and Amazon have severly \textcolor{black}{underestimated} volatility in the model. On a related note, the relative returns distributions (Figures \ref{fig:qof2} j.,k. and l.) show that while the Intel (like Apple) model is able to replicate the fat tailed nature of the empirical returns, the Tesla and Amazon models are not. This suggests that the future work direction should be aimed that solving these problems with high-spread (small-tick) names.

{\footnotesize\section{Disclaimer}}

{\footnotesize Opinions and estimates constitute our judgement as of the date of this Material, are for informational purposes only and are subject to change without notice. This Material is not the product of J.P. Morgan’s Research Department and therefore, has not been prepared in accordance with legal requirements to promote the independence of research, including but not limited to, the prohibition on the dealing ahead of the dissemination of investment research. This Material is not intended as research, a recommendation, advice, offer or solicitation for the purchase or sale of any financial product or service, or to be used in any way for evaluating the merits of participating in any transaction. It is not a research report and is not intended as such. Past performance is not indicative of future results. Please consult your own advisors regarding legal, tax, accounting or any other aspects including suitability implications for your particular circumstances. J.P. Morgan disclaims any responsibility or liability whatsoever for the quality, accuracy or completeness of the information herein, and for any reliance on, or use of this material in any way.

\bibliographystyle{plainnat}
{
\renewcommand{\clearpage}{} 
\bibliography{sample-base}
}

\section*{Disclosure Statement}

No potential conflict of interest was reported by the author(s).

\section*{Funding}

This work was supported by JP Morgan Chase \& Co.

\end{document}